\documentclass[useAMS,usenatbib]{mnras}
\usepackage{multirow}
\usepackage{epsfig}
\usepackage{subfigure}
\usepackage{graphicx}
\usepackage{amsmath,amssymb}

\makeatletter
\let\jnl@style=\rm
\def\ref@jnl#1{{\jnl@style#1}}

\def\aj{\ref@jnl{AJ}}                   
\def\araa{\ref@jnl{ARA\&A}}             
\def\apj{\ref@jnl{ApJ}}                 
\def\apjl{\ref@jnl{ApJ}}                
\def\apjs{\ref@jnl{ApJS}}               
\def\ao{\ref@jnl{Appl.~Opt.}}           
\def\apss{\ref@jnl{Ap\&SS}}             
\def\aap{\ref@jnl{A\&A}}                
\def\aapr{\ref@jnl{A\&A~Rev.}}          
\def\aaps{\ref@jnl{A\&AS}}              
\def\azh{\ref@jnl{AZh}}                 
\def\baas{\ref@jnl{BAAS}}               
\def\jrasc{\ref@jnl{JRASC}}             
\def\memras{\ref@jnl{MmRAS}}            
\def\mnras{\ref@jnl{MNRAS}}             
\def\pra{\ref@jnl{Phys.~Rev.~A}}        
\def\prb{\ref@jnl{Phys.~Rev.~B}}        
\def\prc{\ref@jnl{Phys.~Rev.~C}}        
\def\prd{\ref@jnl{Phys.~Rev.~D}}        
\def\pre{\ref@jnl{Phys.~Rev.~E}}        
\def\prl{\ref@jnl{Phys.~Rev.~Lett.}}    
\def\pasp{\ref@jnl{PASP}}               
\def\pasj{\ref@jnl{PASJ}}               
\def\qjras{\ref@jnl{QJRAS}}             
\def\skytel{\ref@jnl{S\&T}}             
\def\solphys{\ref@jnl{Sol.~Phys.}}      
\def\sovast{\ref@jnl{Soviet~Ast.}}      
\def\ssr{\ref@jnl{Space~Sci.~Rev.}}     
\def\zap{\ref@jnl{ZAp}}                 
\def\nat{\ref@jnl{Nature}}              
\def\iaucirc{\ref@jnl{IAU~Circ.}}       
\def\aplett{\ref@jnl{Astrophys.~Lett.}} 
\def\apspr{\ref@jnl{Astrophys.~Space~Phys.~Res.}}
\def\bain{\ref@jnl{Bull.~Astron.~Inst.~Netherlands}}
\def\fcp{\ref@jnl{Fund.~Cosmic~Phys.}}  
\def\gca{\ref@jnl{Geochim.~Cosmochim.~Acta}}   
\def\grl{\ref@jnl{Geophys.~Res.~Lett.}} 
\def\jcp{\ref@jnl{J.~Chem.~Phys.}}      
\def\jgr{\ref@jnl{J.~Geophys.~Res.}}    
\def\jqsrt{\ref@jnl{J.~Quant.~Spec.~Radiat.~Transf.}}
\def\memsai{\ref@jnl{Mem.~Soc.~Astron.~Italiana}}
\def\nphysa{\ref@jnl{Nucl.~Phys.~A}}   
\def\physrep{\ref@jnl{Phys.~Rep.}}   
\def\physscr{\ref@jnl{Phys.~Scr}}   
\def\planss{\ref@jnl{Planet.~Space~Sci.}}   
\def\procspie{\ref@jnl{Proc.~SPIE}}   

\makeatother

\title[Radiation Pressure Compression in the X-ray Narrow Line Region of Seyfert galaxies]{Evidence for Radiation Pressure Compression in the X-ray Narrow Line Region of Seyfert galaxies}

\author[Stefano Bianchi, et al.]{Stefano Bianchi$^1$\thanks{E-mail: bianchi@fis.uniroma3.it (SB)}, Matteo Guainazzi$^{2}$, Ari Laor$^{3}$, Jonathan Stern$^{4}$, Ehud Behar$^{3}$
\\
$^1$Dipartimento di Matematica e Fisica, Universit\`a degli Studi Roma Tre, via della Vasca Navale 84, 00146 Roma, Italy\\
$^2$ESA/ESTEC, D-SRE, Keplerlaan 1, 2200 AG, Noordwijk, The Netherlands\\
$^3$Physics Department, Technion - Israel Institute of Technology, Haifa 3200000, Israel\\
$^4$Department of Physics and Astronomy and CIERA, North-western University, Evanston, IL, USA\\
}
\begin{document}

\maketitle

\label{firstpage}

\begin{abstract}
The observed spatial and kinematic overlap between soft X-ray emission and the Narrow Line Region (NLR) in obscured Active Galactic Nuclei (AGN) yields compelling evidence that relatively low-density gas co-exists with higher density gas on scales as large as 100s of pc. This is commonly interpreted as evidence for a constant gas pressure 
multiphase medium, likely produced by thermal instability. Alternatively, Radiation Pressure Compression (RPC) also leads to a density distribution, since a gas pressure (and hence density) gradient must arise within each cloud to counteract the incident ionising radiation pressure. RPC leads to a well-defined ionization distribution, and a 
Differential Emission Measure (DEM) distribution with a universal slope of $\sim-0.9$, weakly dependent on the gas properties and the illuminating radiation field. In contrast, a multiphase medium does not predict the form of the DEM. The observed DEMs of obscured AGN with XMM-\textit{Newton} RGS spectra (the CHRESOS sample) are in striking agreement with the predicted RPC DEM, providing a clear signature that RPC is the dominant mechanism for the observed range of densities in the X-ray NLR. In contrast with the constant gas pressure multiphase medium, RPC further predicts an increasing gas pressure with decreasing ionization, which can be tested with future X-ray missions using density diagnostics.
\end{abstract}


\newcommand{\JS}[1]{{\color{red}{#1}}}
\newcommand{\nH}{n_{\rm H}}
\section{Introduction}
\label{introduction}

Soft X-ray emission above the extrapolation of the absorbed nuclear emission is ubiquitous in low resolution spectra of nearby X-ray obscured Seyfert galaxies \citep[e.g.][]{Turner1997,Guainazzi2005}. Whenever high resolution spectra are available, thanks to the gratings aboard \textit{Chandra} and XMM-\textit{Newton}, this emission observed in CCD spectra is resolved into strong emission lines, mainly from He- and H-like transitions of light metals and L-shell transitions of Fe \citep[e.g.][]{Sako2000,Kinkhabwala2002,Schurch2004,Bianchi2005a,Bianchi2010a,Guainazzi2007,Guainazzi2009,Nucita2010,Marinucci2011,Whewell2015,Braito2017}. 
Several pieces of evidence converge towards a scenario where the emitting gas is photoionized by the radiation field of the active nucleus, in particular the clear detection of narrow Radiative Recombination Continua (RRC) features,  typical signature of low-temperature plasma \citep{Liedahl1996}. 

High resolution X-ray imaging added an insightful ingredient to the overall scenario, showing that the soft X-ray emission of obscured Seyfert galaxies is extended on 100s of pc, and strongly correlated with the optical Narrow Line Region (NLR), as mapped by the [{O\,\textsc{iii}}] $\lambda 5007$ emission \citep[e.g.][]{Young2001,Bianchi2006,Bianchi2010a,Levenson2006, Greene2014, Dadina2010,Maksym2017,Fabbiano2018}. 
Since the X-ray emission lines trace gas at a higher ionization level than the optical lines, this observed spatial overlap suggests the relatively low-density gas which produces the X-ray emission co-exists with the higher density gas which produces the optical emission.

Which mechanism can produce such a local density gradient in photoionized gas? 
A classic solution is the `thermal instability' of \citet{Krolik1981} where high-density low-temperature gas is confined by low-density high-temperature gas with the same thermal pressure. 
This model however neglects the effect of radiation pressure on the gas, which is easily the dominant pressure source near the AGN. Specifically, in gas with the ionization parameter of $\xi=10^2-10^4$ erg cm s$^{-1}$ required to produce some of the X-ray emission lines, the pressure in the incident radiation exceeds the thermal gas pressure by a factor of $\sim 100$. This is evident from the definition of $\xi$
\begin{equation}\label{eq: xi}
 \xi \equiv \frac{L}{\nH r^2}
\end{equation}
where $L$ is the AGN luminosity, $\nH$ is the hydrogen density\footnote{We use both the hydrogen density $\nH$ and the electron density $n_e$ where appropriate. However, since hydrogen and helium are fully stripped at the ionization parameters we investigate here, the ratio $\nH/n_e$ is constant, so we may use one or the other sometimes to simplify formulae.} and $r$ is the distance. The ratio of the incident radiation pressure $L/4\pi r^2 c$ to the gas thermal pressure $2\nH k T$ is hence $96\, (T/10^5\,{\rm K})^{-1} (\xi/1000\,\mathrm{erg\, cm\, s}^{-1})$, implying that neglecting the radiation pressure is generally not justified, unless the gas optical depth is extremely optically thin, which is not possible given the observed strength of the X-ray lines. 

As shown by \citet{Stern2014b}, an alternative explanation for the overlap of X-ray and optical emission is the mechanism of Radiation Pressure Compression (RPC) \citep{Dopita2002, Draine2011} where the incident radiation pressure compresses the ionized gas against the shielded side of the cloud. In this scenario, the \ion{H}{ii} layer of the cloud forms a pressure gradient to counteract the force induced by the momentum of the absorbed photons. The result is an \ion{H}{ii} surface layer which spans a huge dynamical range of $\gtrsim 10^4$ in density and ionization, over a physical size much smaller than the distance of the cloud to the source \citep[see Fig.~2 in][]{Stern2014a}. Hence, the X-ray emission originating from the low-density highly-ionized gas will appear co-spatial with the optical emission originating in the high-density gas. 

Another advantage of the RPC mechanism is that the entire structure of the \ion{H}{ii} layer is set by the combination of ionization, thermal and hydrostatic equilibrium equations, and hence the expected emission or absorption emission spectrum can be calculated with a minimal amount of free parameters. Previous papers have compared the predicted spectrum with various observables of ionized gas in AGN spectra, and found that RPC can explain the characteristic ionization parameter in the NLR of Seyferts \citep{Dopita2002} and quasars \citep{Stern2016}, the characteristic ionization of the BLR over a range of $10^8$ in AGN luminosity \citep{Baskin2014a}, the broad ionization distribution of warm absorbers \citep{Rozanska2006, Stern2014a}, and the small filling factor of broad absorption lines \citep{Baskin2014b}. 
The goal of this paper is to further test the predictions of RPC against observed X-ray line emission in obscured Seyferts. As we show below, X-ray emission lines originate from gas over a large range of ionization levels ($30<\xi<3000$ erg cm s$^{-1}$), and hence the X-ray spectra are best-suited to test the RPC mechanism.

\section{Radiation Pressure Compression}
\label{rpcintro}
\newcommand{\sigmarp}{\tilde{\sigma}}

Consider an optically thick cloud of gas at a distance $r$ from the irradiating source, where the size of the cloud is $\ll r$ so the incident radiation can be considered as plane-parallel. Assume that the incident radiation is absorbed by the cloud, while the radiation emitted from the gas escapes the system without further interaction. If the total column of the cloud is high enough so gravity dominates radiation pressure (e.g.\ $>10^{24}\,{\rm cm}^{-2}$ for Eddington luminosity), the gas remains confined and the illuminated surface layer gets compressed into the back side. The pressure gradient in the illuminated surface is hence:
\begin{equation}\label{e:hydrostatic}
 \frac{d P_{\rm gas}}{d r} = \frac{F(r)}{c} n_{\rm H} \sigmarp ~~~,
\end{equation}
where $P_{\rm gas}$ is the gas thermal pressure, $F(r)$ is the flux in the incident radiation which has not been absorbed at $<r$, and $\sigmarp$ is the spectrum-averaged cross section per H-nucleon:
\begin{equation}\label{eq: sigbar definition}
 \sigmarp\equiv\frac{\int \sigma_{\nu} F_\nu{\rm d}\nu }{\int F_\nu{\rm d}\nu } ~~~,
\end{equation}
where $F_\nu$ is the flux density and $\sigma_{\nu}$ is the absorption cross-section per H-nucleon. Scattering and magnetic pressure are neglected. We can simplify eqn.~(\ref{e:hydrostatic}) using the definition of the flux-averaged optical depth
\begin{equation}
{\rm d}\tilde{\tau} = \nH\sigmarp {\rm d} r ~~~,
\end{equation}
which implies
\begin{equation}\label{eq: hydro simple form}
 \frac{{\rm d}P_{\rm gas}}{{\rm d}\tilde\tau} = \frac{F(r)}{c} ~~~.
\end{equation}
Using also
\begin{equation}
 F = \frac{L}{4\pi r^2} e^{-\tilde\tau}
\end{equation}
and neglecting the small geometric dilution of the radiation within the cloud (i.e. $F$ changes only via absorption), we get the solution 
\begin{equation}\label{eq: hydro solution 0}
 P_{\rm gas}(\tilde\tau) = P_{\rm gas;0} + \frac{L}{4\pi r^2 c}\left(1-e^{-\tilde\tau}\right) ~~~.
\end{equation}
The constant of integration $P_{\rm gas;0}$ is the gas pressure at the illuminated surface of the cloud, which is set by external pressure sources other than radiation, such as a shocked wind bubble or the ram pressure of disk winds. We assume that in all layers of interest these alternative pressure sources are subdominant to radiation pressure, and hence $P_{\rm gas;0}$ can be neglected. Therefore, in RPC
\begin{equation}\label{eq: hydro solution}
 P_{\rm gas}(\tilde\tau) = \frac{L}{4\pi r^2 c}\left(1-e^{-\tilde\tau}\right) ~~~.
\end{equation}
This equation states that in RPC clouds the gas pressure in some layer equals the momentum of the radiation that has been absorbed from the illuminated surface up to that layer.

Below, we use {\sc cloudy} to numerically solve equation~(\ref{eq: hydro solution}), coupled with the equations for thermal and ionization balance which determine $T$ and $\tilde\sigma$ as a function of depth into the cloud. The {\sc cloudy} calculation hence gives the structure and emission properties of an RPC cloud. We first though derive some analytic approximations of the solution. 
In layers where $\tilde\tau\gtrsim 1$, we can neglect the $e^{-\tilde\tau}$ term in Equation~(\ref{eq: hydro solution}) and we get
\begin{equation}
 P_{\rm gas}(\tilde{\tau}\gtrsim  1) \approx \frac{L}{4\pi r^2 c} ~.
\end{equation}
Using $P_{\rm gas}=2\nH kT$ and the definition of $\xi$ (eqn.~\ref{eq: xi}) we get
\begin{equation}
 \xi(\tilde{\tau}\gtrsim 1)  = 8\pi k T c= 1.04\,T_4  ~.
\end{equation}
where $T=10^4 T_4\,{\rm K}$ and $\xi$ is in cgs units. This relation implies that in optically thick RPC clouds most of the absorption (and hence emission) is in a layer with $\xi\approx 1$ erg cm s$^{-1}$, as derived by \cite{Dopita2002}. 
On the other hand, in layers with $\tilde{\tau}\ll1$ we can approximate $1-e^{-\tilde\tau}\approx \tilde\tau$, so eqn.~(\ref{eq: hydro solution}) implies 
\begin{equation}
 2\nH k T = \frac{L}{4\pi r^2 c}\tilde\tau
\end{equation}
or
\begin{equation}\label{eq: tau<<1}
 \xi(\tilde{\tau}\ll 1) = 8\pi k T c \tilde{\tau}^{-1} = 1.04\,T_4\tilde{\tau}^{-1} ~.
\end{equation}
In practice, thermal balance in photoionized media dictates that $T$ increases with $\xi$. A rough approximation for dust-less gas with $1<\xi<3000$ erg cm s$^{-1}$, solar metallicity, and an ionizing spectral slope of $-1.6$ is \citep[see][]{Stern2014a}
\begin{equation}\label{e:T power-law}
 T(\xi) \approx 10^{4.2}\, \xi^{0.5}\,{\rm K} ~,
\end{equation}
so we get 
\begin{equation}\label{e:xi approx}
 \xi(\tilde{\tau}\ll 1) \approx 2.7 \,\tilde\tau^{-2}
\end{equation}
In an RPC cloud, the low-$\tilde\tau$ high-$\xi$ surface layers characterised by eqn.~(\ref{e:xi approx}) produce the X-ray emission lines which are the focus of this paper.

\subsection{The Emission Measure Distribution}
\label{emd}

The luminosity of an emission line of a given ion can be expressed as the volume integral of its emissivity $j_{ul}(\xi)$ at the ionization parameter $\xi$ of the emitting gas
\begin{equation}
L_{ul}=\int_V j_{ul}(\xi) dV
\label{luminosity}
\end{equation}
If we define the \textit{emission measure} as $\mathrm{EM}=\int_V n_e^2 dV$ and the line power $P_{ul}(\xi)=j_{ul}/n_e^2$, we can rewrite (\ref{luminosity}) in terms of  a differential line luminosity produced over a range of ionization parameters $d \log\xi$
\begin{equation}
L=\int_\xi d \log\xi \left[ \frac{d\left(\mathrm{EM}\right)}{d\log \xi}  \right] P_{ul}(\xi)
\label{luminositydem}
\end{equation}
The bracketed quantity above represents the \textit{differential emission measure (DEM) distribution} \citep[e.g.][]{Liedahl1999,Sako1999}:
\begin{equation}
 \frac{d \left(\mathrm{EM}\right)}{d\log \xi} 
 = n_e^2(\xi) \frac{d V}{d\log \xi}
\end{equation}
 In practice, the DEM distribution is the ensemble of weighting factors that determine the contributions of each ionization zone to the total line flux. This is conceptually the same as for the Absorption Measure Distribution \citep[AMD, e.g.][]{Holczer2007}, used for describing the absorbing properties of photoionized gas. An analogous quantity is widely used in the context of  plasmas in collisional equilibrium ($d\left(\mathrm{EM}\right)/d \log T$). In this context, it is customary to convert the integral (\ref{luminositydem}) into a sum over finite bins \citep[e.g.][]{Kaastra1996}. This is the same approach we will follow in the detailed methodology described below.

 The usefulness of the DEM distribution is that it can be derived theoretically for a given scenario of X-ray emission, and then can be readily compared to what is measured experimentally, via spectroscopic analysis. An emblematic example is the case of gas with constant density illuminated by a point source, which results in the characteristic $d \left(\mathrm{EM}\right)/d  \log\xi \propto \xi^{-3/2}$ \citep[e.g.][]{Liedahl1999}. In general, it was shown that a power-law density profile produces a power-law DEM curve, a constant flat DEM being the specific result of a radial density profile $n_e(r)\propto r^{-3/2}$ \citep{Sako2000}.

\subsection{The DEM in RPC}
\label{rpcdem}

What is the DEM expected in RPC? Since the RPC structure is plane parallel we can replace $d V$ with $\Omega r^2 d r$, where $\Omega$ is the solid angle subtended by the illuminated clouds. So
\begin{equation}\label{e:DEM analytic}
 \frac{d \left(\mathrm{EM}\right)}{d\log \xi} 
  = \Omega \nH^2r^2\left|\frac{d r}{d\log \xi}\right|
\end{equation}
An approximation of $|d r/d\log \xi|$, the \textit{ionization scale length}, can be derived by rearranging eqn.~(\ref{eq: tau<<1}) 
\begin{equation}
 \frac{8\pi kT c}{\xi} = \tilde\tau = \int \nH\sigmarp d r
\end{equation}
Differentiating with respect to $r$ and dividing by $T/\xi$ gives
\begin{equation}
 \frac{d \log T/\xi}{d r} = \frac{\nH\xi\sigmarp}{8\pi k Tc}
\end{equation}
and using $d\log T/\xi = Ad\log \xi$ with $A\equiv d \log T/d \log \xi-1$ we hence get
\begin{equation}\label{e:l xi}
 \left|\frac{d r}{d \log \xi}\right| = |A|\frac{8\pi k Tc r^2}{L\sigmarp}
\end{equation}
where we replaced $\nH$ with $L/\xi r^2$. 
The DEM can then be derived by using eqn.~(\ref{e:l xi}) in eqn.~(\ref{e:DEM analytic}), which gives
\begin{equation}\label{e:DEM RPC}
 \frac{d \left(\mathrm{EM}\right)}{d\log \xi} = |A|\frac{ 8 \pi  k T c}{\xi^2\sigmarp } \Omega L
\end{equation}
The DEM is proportional to $\Omega L$, as expected for optically thick clouds. Also, note that if $T$ and $\sigmarp$ are only functions of $\xi$, and not directly of $r$, then the DEM is independent of $r$. The DEM of an ensemble of clouds with different $r$ is hence the same as the DEM of a single cloud, and one does not need to resolve the clouds in order to predict their DEM.

To derive numerical estimates for eqns.~(\ref{e:l xi})--(\ref{e:DEM RPC}), we can use eqn.~(\ref{e:T power-law}) for $T(\xi)$  (which also implies $|A|=0.5$), and a similarly derived power-law approximation for $\sigmarp(\xi)$ \citep[see][]{Stern2014a}:
\begin{equation}
 \sigmarp(\xi) \approx 10^{-22}\, \xi^{-0.6}\, {\rm cm}^{-2} ~.\\
\end{equation}
The ionization scale length is hence
\begin{equation}\label{e:l xi num}
 \left|\frac{d r}{d \log \xi}\right| = 0.021 \, L_{X; 43}^{-1} r_{\rm pc}^{2} \xi_{100}^{1.1}\, {\rm pc}
\end{equation}
while the DEM is
\begin{equation}\label{e:DEM RPC num}
 \frac{d \left(\mathrm{EM}\right)}{d\log \xi} =  3.1 \cdot10^{65}\, \Omega_{4\pi} L_{X;43} \xi_{100}^{-0.9} \,{\rm cm}^{-3}
\end{equation}
where we defined $\Omega_{4\pi}=\Omega/4\pi$, $r_{\rm pc}=r/{\rm pc}$, $\xi_{100}=\xi/100$ erg cm s$^{-1}$, and $L_{X;43}$ corresponds to a $2-10$ keV X-ray luminosity of $10^{43}$ erg s$^{-1}$, which we convert to $L$ using a bolometric correction of 19, an average over the $10^{42}-10^{44}$ erg s$^{-1}$ luminosity range of local Seyfert galaxies (\citealt{Marconi2004}). 
Thus, in dust-less RPC clouds we expect a differential emission measure which decreases roughly inversely with $\xi$. 
This result can be compared to the flat ($\propto \xi^0$) AMD expected in RPC \citep{Stern2014a}. Since the AMD is $\propto \nH d r$ while the DEM is $\propto \nH^2 d r$, and $\xi\propto \nH^{-1}$, the index of the DEM is one lower than the index of the AMD. For a dusty gas $\sigmarp$ is constant, leading to a steeper DEM, as already noted for the AMD \citep{Stern2014a}.

A more accurate calculation of the DEM in the case of a RPC gas can be derived with \textsc{Cloudy}. We use \textsc{Cloudy} 17.00 \citep[last described in][]{Ferland2013}, adopting the \textsc{constant total pressure} flag \citep{Pellegrini2007}, which increases gas pressure between consecutive zones, according to the attenuation of the incident continuum \citep[eqn.~\ref{e:hydrostatic}, see also][]{Stern2014a,Stern2014b,Stern2016,Baskin2014a,Baskin2014b}. We assume a slab with $\log \mathrm{N_H/cm^{-2}}=23$) , enough to reach the ionization front, a ionization parameter at the illuminated face $\log \xi/\mathrm{erg\,cm\, s^{-1}}=4$, high enough for the ions investigated here, an initial electron density $n_e=1$ cm$^{-3}$, the default \textsc{Cloudy} Solar abundances, illuminated by a generic AGN Spectral Energy Distribution (SED) similar to the one used by \citet{Baskin2014a} and \citet{Stern2014a}, and described in the following. In the range between 1 $\mu$m and 1 Ryd, the SED is described by the expression

\begin{equation}
f_\nu = \nu^{\alpha_\mathrm{UV}} \exp(-h \nu/kT_\mathrm{BB}) \exp(-kT_\mathrm{IR}/h \nu)
\end{equation}

with $\alpha_\mathrm{UV}=−0.5$, $kT_\mathrm{BB}=13$ eV and $kT_\mathrm{IR}=0.1$ eV. Between 1 Ryd and 2 keV, the SED is a simple power-law with $\alpha_{ion}=-1.6$, while above 2 keV the slope becomes $\alpha_x=-0.9$. A cut-off is assumed for $\lambda>1 \mu$m and above 100 keV.

\begin{figure*}
\includegraphics[width=\columnwidth]{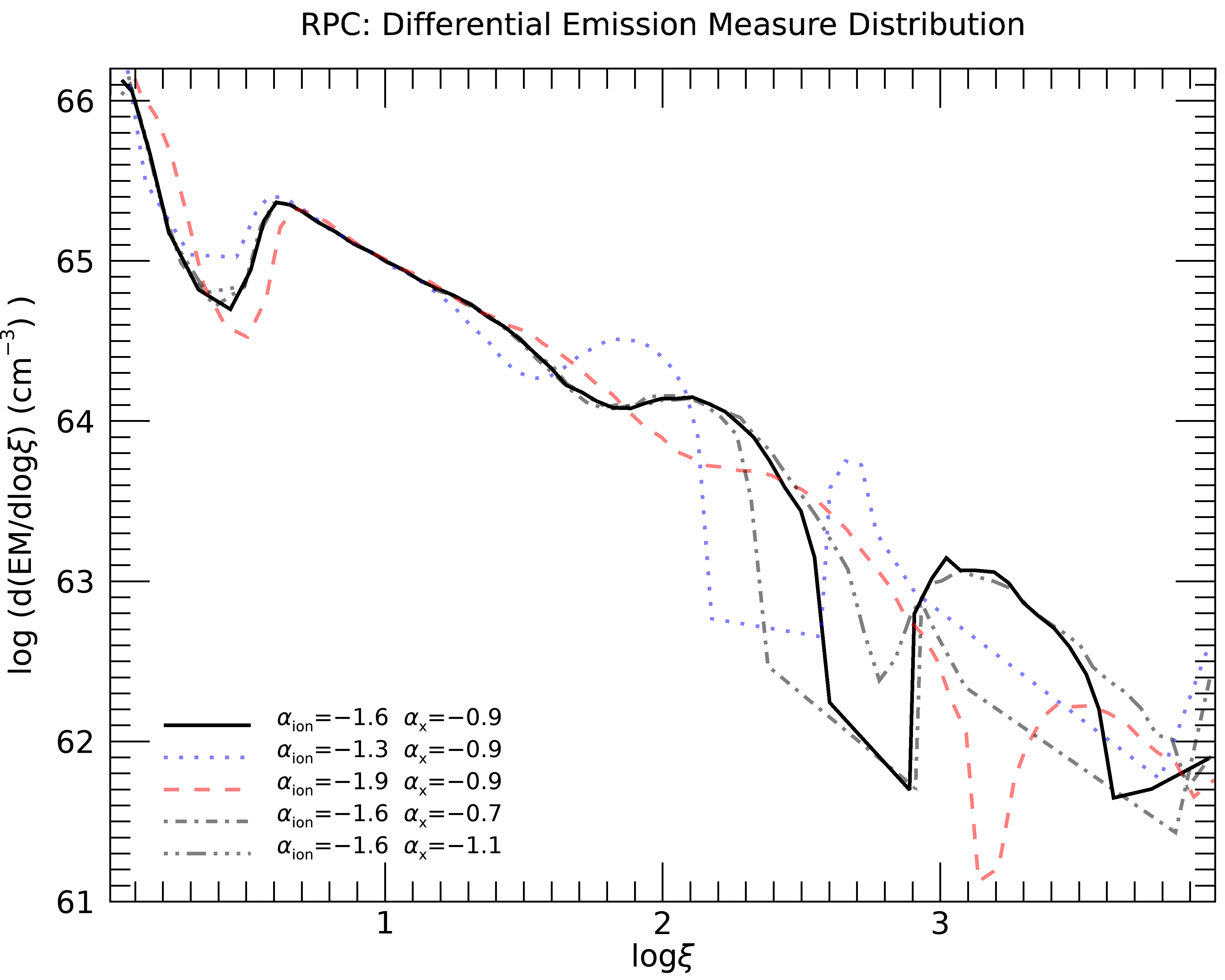}
\includegraphics[width=\columnwidth]{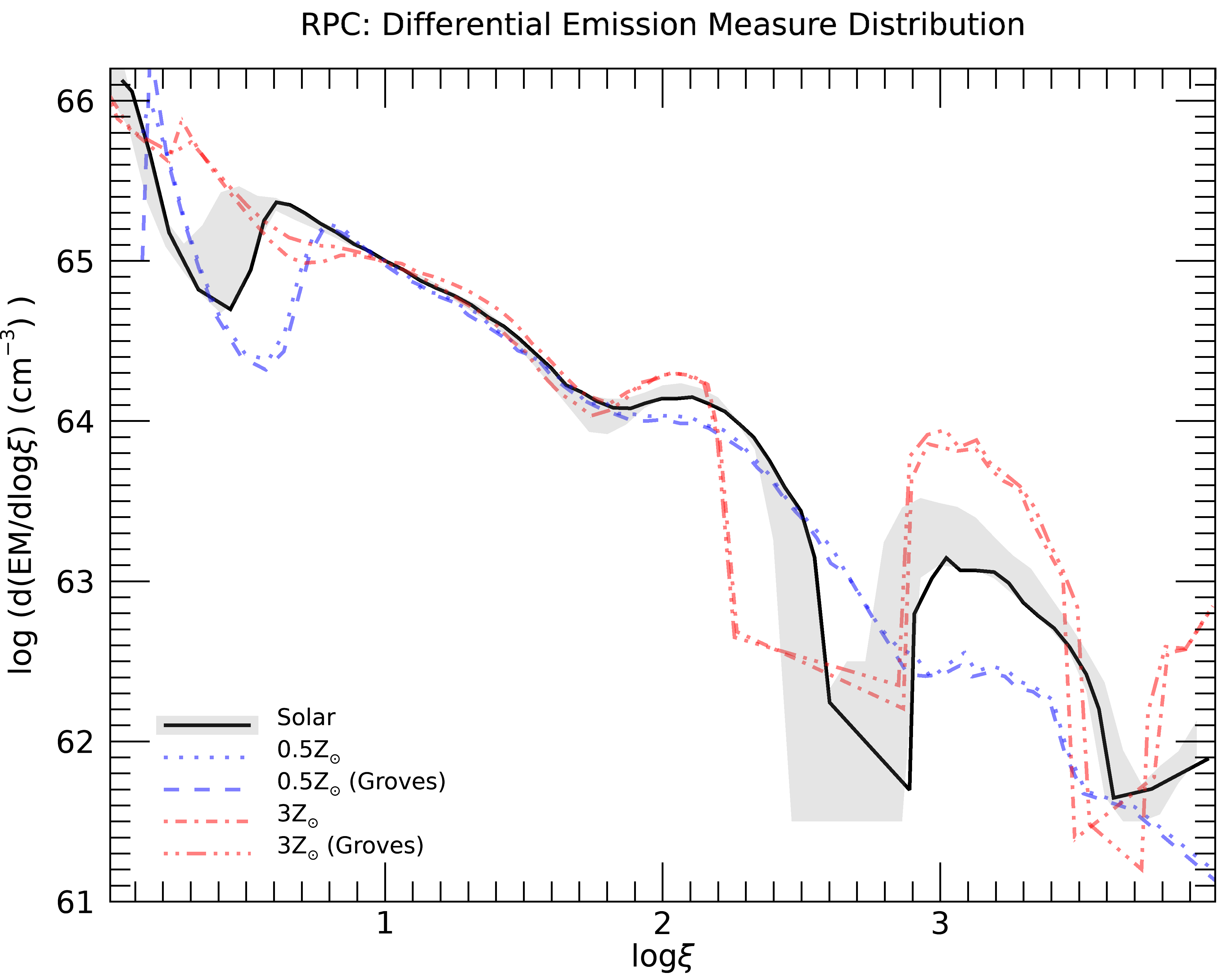}
\caption{\label{EMD_RPC}The Differential Emission Measure Distribution for a Radiation Pressure Compressed gas. The overall normalization is arbitrary. \textit{Left}: The black solid curve is for the default SED adopted in this paper, while the other curves are derived with different values of $\alpha_{ion}$ and $\alpha_x$, as reported in the legend. \textit{Right}: The grey shaded area encompasses the differences introduced by different sets of Solar abundances, with respect to the default used in this work (black curve). Red curves are the DEM distributions for a metal over-abundance by a factor of 3, while the blue curves for a metal under-abundance by a factor of 0.5, with or without the \ion{N}{}/\ion{O}{} prescription in \citet{Groves2004} (see text for details).}
\end{figure*}

The corresponding DEM is directly derived as $d \left(\mathrm{EM}\right)/d  \log\xi$, by using the relevant quantities in each zone of the slab computed by \textsc{Cloudy} (see Sect.~\ref{emd}). The result is shown in the left panel of Fig.~\ref{EMD_RPC} (black solid line): the derived DEM has an approximate power-law shape, with a slope of $-0.99$ (in the range $\log\xi/\mathrm{erg\,cm\, s^{-1}}=0:4$), similar to the analytic approximation in eqn.~\ref{e:DEM RPC num}, and much flatter than the expected $-1.5$ arising in a constant-density slab \citep[e.g.][]{Liedahl1999}.

Superimposed over the overall power-law shape, there are few `troughs' at specific ranges of ionization parameters (around $\log\xi/\mathrm{erg\,cm\, s^{-1}}=0.4, 2.8, 3.7$). They are likely due to thermal instabilities. Similar troughs are observed in the absorption measure distributions of warm absorbers \citep[e.g.][]{Holczer2007,Behar2009,Adhikari2015,Goosmann2016a}.

We then investigated how the DEM curve shape is robust against the specific parameters of the illuminating SED and the gas properties. The dependency of the RPC solutions upon column density and ionization parameter have been extensively discussed in previous papers \citep[e.g.][]{Stern2014a,Stern2014b,Stern2016,Baskin2014a,Baskin2014b}. The value of the ionization parameter at the illuminated face only sets the highest ionization considered for our DEM. Adopting a value larger than $\log \xi/\mathrm{erg\,cm\, s^{-1}}=4$ would only add a fully ionized layer at the front of our cloud, which would be effectively transparent for the production of the recombination lines treated here. On the other hand, a lower value would introduce a cut-off in the DEM at higher ionization parameters. However, this would be physically justified only by introducing another process which compresses the gas, in order to suppress the otherwise unavoidable ionized layer of each RPC solution (see Sect.~\ref{discussion}).

In a similar fashion, a column density higher than $\log \mathrm{N_H}/\mathrm{cm^{-2}}=23$ would not change the derived DEM in the $\log \xi$ range investigated here, because it would only add emission at even lower ionization parameters. On the other hand, a lower column density would flatten the DEM, because the cloud would begin to lack the lower ionization layers responsible for the low $\log \xi$ part of the curve. Low column density clouds may certainly exist, and their effect on the observed DEMs will be discussed in Sect.~\ref{discussion}.

No dependence upon electron density is expected, because all solutions will be self-similar with respect to this parameter, provided it is lower than the critical density of the forbidden transitions considered here \citep[from $5.3\times10^9$ to $8.6\times10^{13}$ cm$^{-3}$ for \ion{N}{vi} and \ion{Si}{xiii}, respectively: ][]{Gudel2009}. However, \citet{Adhikari2015} showed that the $\xi$ ranges where thermal instabilities occur do depend on density. Similarly to what found by \citet{Adhikari2015}, we find that the dependence on density is significant only for very high densities ($n_e>10^{10}$ cm$^{-3}$), which are more typical for the Broad Line Region, and inconsistent with the observed forbidden lines in the NLR.

The dependence of the DEM distribution upon the illuminating SED is more subtle, and deserves a deeper investigation. We therefore varied the default SED described above. In particular, along with the default $\alpha_{ion}=-1.6$, we tested $\alpha_{ion}=-1.3$, $-1.9$, corresponding to optical to X-ray slopes in the range $\alpha_{ox}=[-1.3:-1.8]$, as observed \citep[e.g.][]{Steffen2006,Just2007}. Moreover, we varied the X-ray slope from the default $\alpha_x=-0.9$ to $\alpha_x=-0.7$, $-1.1$, to reflect the X-ray spectral index distribution observed in local Seyfert galaxies \citep[e.g.][]{Bianchi2009}. The results are shown in the left panel of Fig.~\ref{EMD_RPC}. Overall, the DEM is not strongly affected by the illuminating SED, becoming slightly steeper (slope of $-1.12$) for steeper ionizing slopes $\alpha_{ion}$, and slightly flatter ($-0.95$) for flatter $\alpha_{ion}$. The positions of the unstable troughs are, instead, significantly different for different $\alpha_{ion}$. The effect of a variation of the X-ray slope is less important, being almost irrelevant in the case of a steeper $\alpha_x$ (same DEM slope of $-0.99$, and very similar unstable troughs), somewhat more important for a flatter $\alpha_x$ (steeper DEM slope of $-1.06$, larger unstable troughs).

We next investigated the effect of chemical abundances on the derived DEM. As a first step, we looked for the effect of the uncertainties on the standard Solar abundance. We therefore derived the DEM for seven different Solar abundance data-sets, i.e. the ones implemented in \textsc{Xspec} \citep[last described in][]{Arnaud1996}: \citet{Anders1982,Anders1989,Feldman1992,Grevesse1998,Wilms2000,Lodders2003,Asplund2009}. The result is shown in the right panel of Fig.~\ref{EMD_RPC} (grey shaded area). The different Solar abundances do not introduce any relevant effect on the overall shape of the derived DEM, but can change the width and depth of the unstable troughs. Finally, we investigated the effect of metallicity. At first we simply multiply the abundance of all elements above \ion{He}{} by constant factors of $0.5$ and $3$. The effects of these variations of metallicity are shown in Fig.~\ref{EMD_RPC} (red and blue curves): again, their main effect is in the unstable troughs. In particular, it is notable that metal under-abundance almost washes out the trough at $\log\xi/\mathrm{erg\,cm\, s^{-1}}\sim2.8$. As a further step, we considered the peculiar case of nitrogen, which is known to possess both a primary and secondary nucleosynthetic component. We therefore took into account the \citet{Groves2004} prescriptions on the \ion{N}{}/\ion{O}{} ratio, treating it as a rising function of metallicity. The results are again shown in the right panel of Fig.~\ref{EMD_RPC}: the differences with respect to the previous curves are negligible.

Overall, we have shown that the derived DEM distribution in the case of RPC gas is very characteristic and robust against the specific gas parameters and illuminating SEDs. In practice, the DEM is basically set by the hydrostatic equilibrium which the gas must obey in case of RPC (eqn.~\ref{e:hydrostatic}), and does not depend on the other details. 

The next step is to compare these theoretical predictions with the observations. A DEM $\sim \xi^{-1}$ is reported for NGC~1068 by \citet{Ogle2003}, in very good agreement with the RPC expectations. A flat DEM is instead reported for Circinus by \citet{Sako2000}. However, a systematic DEM derivation for observed X-ray spectra of local Seyfert galaxies is still lacking. In the next Section, we will therefore compare the RPC predictions with the DEM distributions derived from the best sample available at the moment, in order to understand if RPC is indeed the universal compressing mechanism of the (X-ray) NLR in these sources.

\section{The CHRESOS sample}
\label{sample}

CHRESOS (a Catalogue of High REsolution Spectra of Obscured Sources) is an extension of the sample discussed in
\cite{Guainazzi2007}. It is based on the analysis of 239 \textit{XMM-Newton} observations of 100 {\it bona fide} X-ray obscured AGN extracted from the \textit{XMM-Newton} Science Archive (XSA). In this paper a total column density of
1$\times$10$^{23}$~cm$^{-2}$ covering the primary nuclear emission separates X-ray ``unobscured'' from X-ray ``obscured'' objects, as we are interested in measuring the X-ray spectra associated to X-ray Extended Narrow Line Regions with the Reflection Grating Spectrometer \citep[RGS:][]{DenHerder2001}, sensitive in the 0.2--2~keV energy
band. The CHRESOS sample is not complete or unbiased in any sense. It represents merely the collection of all good-quality spectroscopic data available in the XSA at the time this study was performed.

For each observation, data were extracted from the archive as raw telemetry (Observation Data Files), and reprocessed
to generate calibrated event lists, source and background spectra, and instrumental responses per observation through the data reduction meta-task {\tt rgsproc} \citep{Gabriel2004}. Source spectra were extracted from a stripe along the dispersion plane whose width along the cross-dispersion direction corresponds to 95\% of the energy-averaged Point Spread Function (PSF). Background spectra were extracted with two methods: from the dispersed CCD outside an area corresponding to the 98\% of the cross-dispersion PSF, and from blank-fields rescaled to the level of quiescent background measured during the observation. Background spectra extracted with the latter method are adopted in this paper. We checked that the results yielded by the two background estimates are consistent within the statistical uncertainties. Intervals of high-level of particle flaring were removed by applying a threshold of 2 counts per
second on a 10-s binned light curve extracted from the peripheral CCD9 \citep{DeVries2015}.

Spectra were analysed with \textsc{xspec 12.9.1} \citep{Arnaud1996}. We derived basic observables (centroid energy, width, intensity, fluxes and luminosities) for a set of emission lines, selected among the strongest bound-bound transitions observed in obscured Seyfert galaxies in the soft X-ray energy band covered by the Reflection Grating Spectrometer on-board \textit{XMM-Newton} \citep[e.g.][]{Kinkhabwala2002, Kallman2014}. These include H- and He-like $\alpha$ transitions from \ion{C}{} to \ion{Si}{}. For He-like triplets, we chose the forbidden transition, which is less affected by line transfer issues, and is generally the strongest line of the triplet \citep[e.g.][]{Guainazzi2007}. We also include the \ion{Fe}{xvii} 3C line, which is the strongest \ion{Fe}{} L line with no blending issues with neighbouring lines. These lines and their laboratory energies are listed in Table~\ref{tab1}.
\begin{table}
  \caption{List of emission line probed in CHRESOS. Ritz wavelengths and energies are from NIST \citep{Kramida2018}, except for \ion{Fe}{xvii} 3C \citep{Brown1998}. The associated ionization parameter bin used for the DEMs is also reported (see text and Fig~\ref{EMD_RPC_lines} for details).}
  \label{tab1}
  \begin{tabular}{l|l|c|c} \hline \hline
$\lambda$ (\AA) & Energy (keV) & Transition & $\log\xi/\mathrm{erg\,cm\, s^{-1}}$\\
\hline
6.182 & 2.0056 & \ion{Si}{xiv} Ly$\alpha$ & $3.57\pm0.17$\\
6.740 & 1.8395 & \ion{Si}{xiii} He$\alpha$ (f)& $2.93\pm0.18$\\
8.421 & 1.4723 & \ion{Mg}{xii} Ly$\alpha$ & $3.42\pm0.23$\\
9.314 & 1.3312 & \ion{Mg}{xi} He$\alpha$ (f) & $2.77\pm0.15$\\
12.133 & 1.0219 & \ion{Ne}{x} Ly$\alpha$ & $3.33\pm0.28$\\
13.699 & 0.9051 & \ion{Ne}{ix} He$\alpha$ (f) & $2.37\pm0.19$\\
15.014 & 0.8258 & \ion{Fe}{xvii} 3C & $2.75\pm0.15$\\
18.969 & 0.6536 & \ion{O}{viii} Ly$\alpha$ & $3.35\pm0.26$\\
22.101 & 0.5610 & \ion{O}{vii} He$\alpha$ (f) & $1.75\pm0.20$\\
24.781 & 0.5003 & \ion{N}{vii} Ly$\alpha$ & $2.61\pm0.20$\\
29.534 & 0.4198 & \ion{N}{vi} He$\alpha$ (f) & $1.45\pm0.21$\\
33.736 & 0.3675 & \ion{C}{vi} Ly$\alpha$  & $2.63\pm0.21$\\
\hline 
\hline
\end{tabular}
\end{table}
Spectral fits were performed using the Cash statistics \citep{Cash1979} that is the maximum likelihood for the Poissonian distribution. Spectra in energy ranges $\pm$40~eV around the nominal observed energy of each transition
were fit with a power-law continuum modified by photo-electric absorption with a column density equal to that due to matter in our Galaxy along the line-of-sight to the corresponding source \citep{Kalberla2005}. Each line was fit with a Gaussian profile leaving all its parameters free in the fit. However, the centroid energy and the line width were
fixed to the (redshift-corrected) laboratory energy and to the instrumental resolution, respectively, if the best-fit values were consistent with them within the statistical uncertainties. We consider a line detected if it yields an improvement in the quality of the fit at the 3$\sigma$ confidence level for one interesting parameter \citep{Lampton1976}. Errors on the line intensity and luminosity are quoted at the 1$\sigma$ level, unless otherwise specified. The adopted cosmological parameters are $H_0=70$ km s$^{-1}$ Mpc$^{-1}$ , $\Omega_\Lambda=0.73$ and $\Omega_m=0.27$.

\subsection{The DEM of CHRESOS}

\begin{figure*}
\includegraphics[width=\columnwidth]{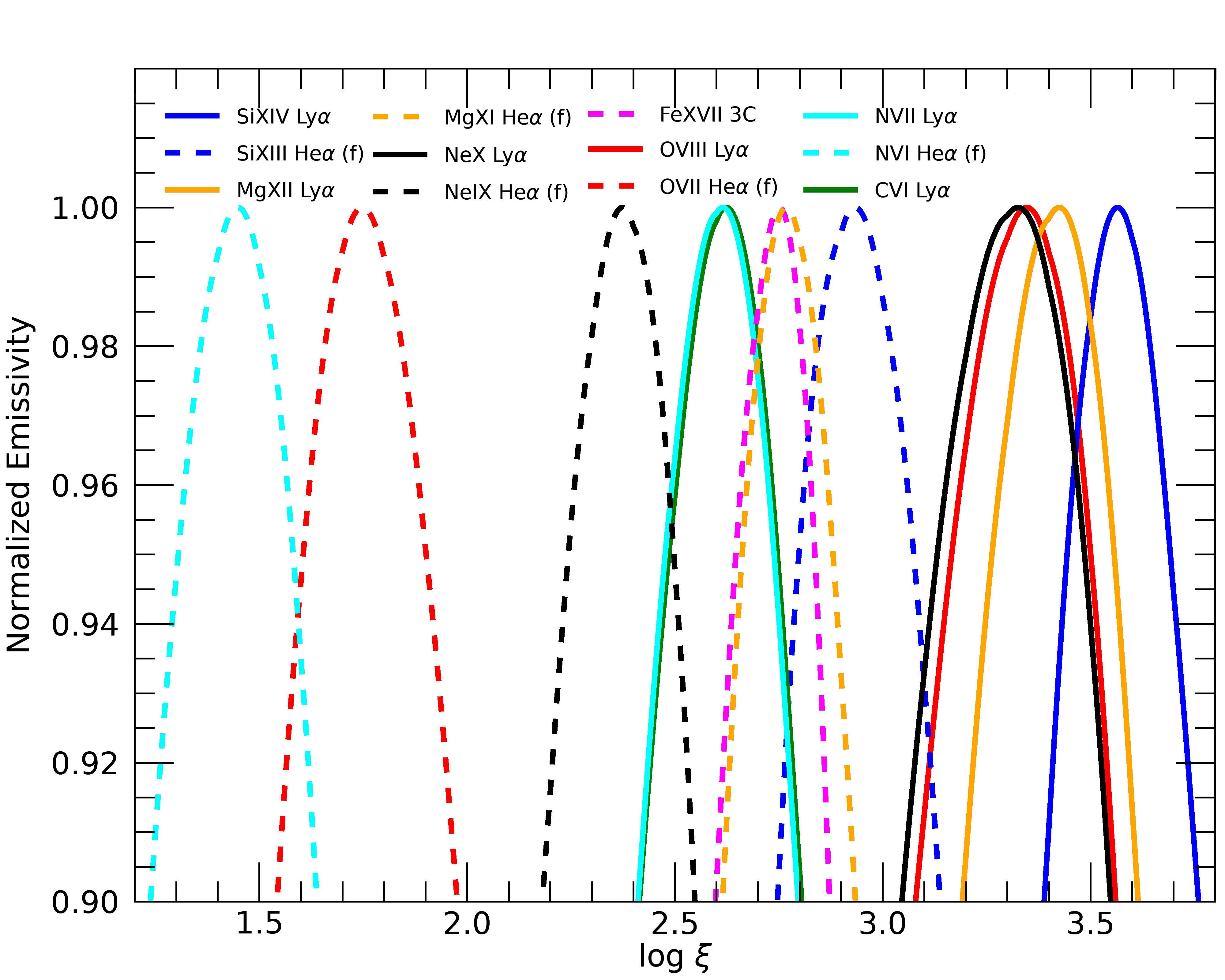}
\includegraphics[width=\columnwidth]{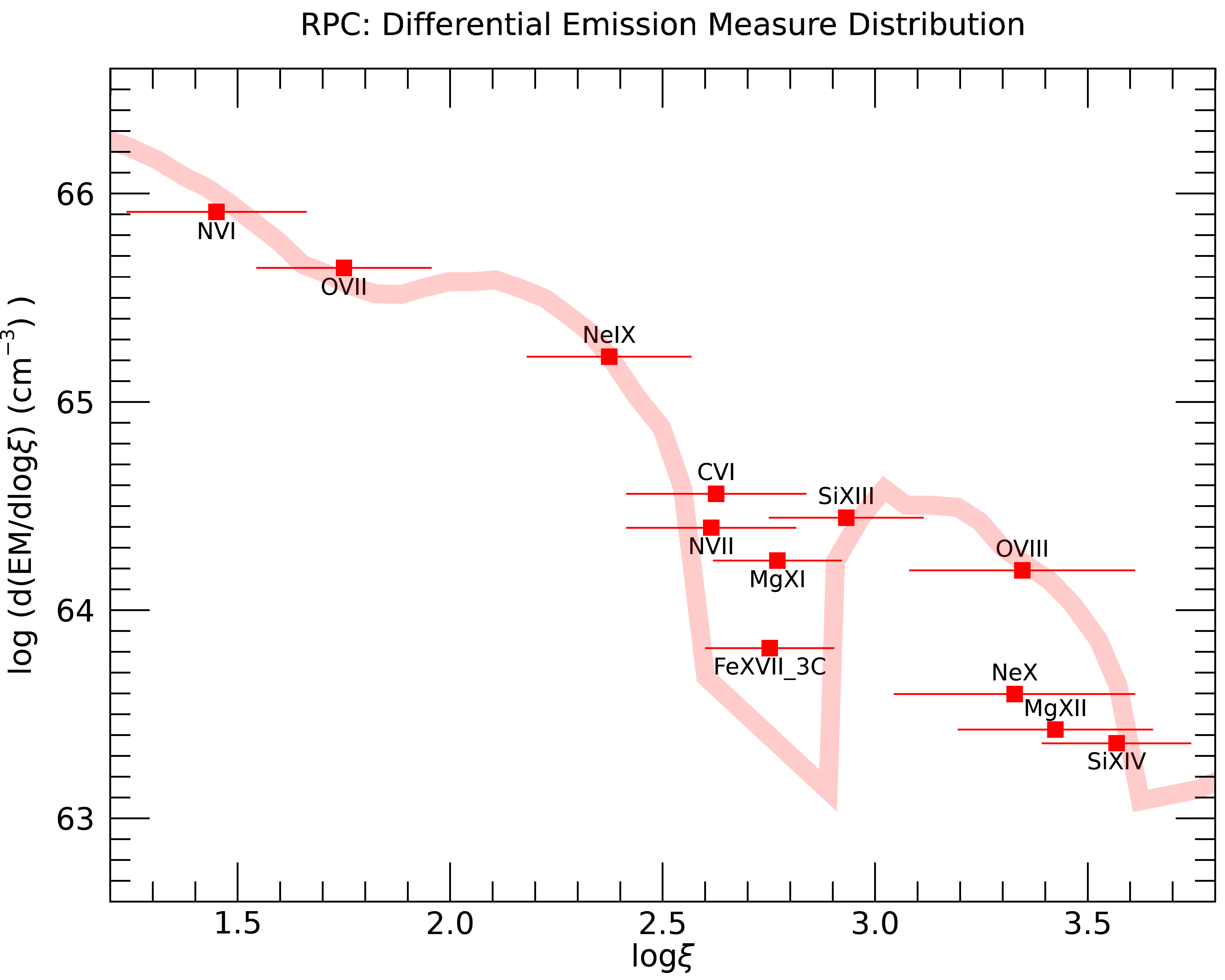}
\caption{\label{EMD_RPC_lines}\textit{Left}: Emissivities for the emission lines listed in Table~\ref{tab1}: these curves are used to associate a ionization parameter to each line (see text for detail) \textit{Right}: A test of our DEM reconstruction methodology for a simulated RPC gas, estimated by using the predicted EMs of the emission lines listed in Table~\ref{tab1} and the associated ionization parameters. The red curve is the intrinsic DEM as derived in Sect.~\ref{rpcdem} and plotted in Fig.~\ref{EMD_RPC}. The overall normalization is arbitrary }
\end{figure*}

The DEM for the observed X-ray spectra can be derived indirectly from the detected emission lines. In this paper, we adopt the following steps:

\begin{itemize}

\item [-] We associate to each emission line in Table~\ref{tab1} a ionization parameter $\log \xi$ and a range $\Delta \log \xi$, corresponding to the peak of its emissivity and its width at 90\% of the peak\footnote{We verified that the results are not affected by the arbitrary choice of the emissivity width, as long as it is self-consistently adopted for all the ions} (left panel of Fig.~\ref{EMD_RPC_lines}). The emissivities were calculated by means of \textsc{Cloudy} computations. 

\item [-] The observed luminosity of each emission line in Table~\ref{tab1} is divided by the corresponding peak emissivity derived above to obtain its EM, once corrected by $n_e^2$ (which is unity in our calculations).

\item [-] The EM of each emission line is divided by the $\Delta \log \xi$ determined above and associated to the corresponding $\log \xi$. The resulting $d \left(\mathrm{EM}\right)/d  \log\xi$ bins constitute the DEM.
\end{itemize}


To verify this methodology, we used it on the RPC computation described in Sect.~\ref{rpcdem}, by using the predicted luminosities for the lines in Table~\ref{tab1}. The result is shown in the right panel of Fig.~\ref{EMD_RPC_lines}: the intrinsic DEM is well sampled by these lines. A linear fit of the 12 points give a slope of $-1.19$, steeper than the intrinsic slope of -0.99 found in Sect.~\ref{rpcdem}. However, the latter, when estimated in the restricted $\log \xi$ range spanned by the chosen emission lines, is $-1.09$. It is important to note that the most significant unstable trough in this range (around $\log \xi/\mathrm{erg\,cm\, s^{-1}}=2.8$) is somewhat washed out by the rougher sampling of the emission lines. With these caveats in mind, we conclude that the DEMs derived with this methodology are fair estimates of the intrinsic DEMs, and we proceeded in using it to derive the DEMs of the CHRESOS sample.

\begin{figure*}
\includegraphics[width=0.48\textwidth]{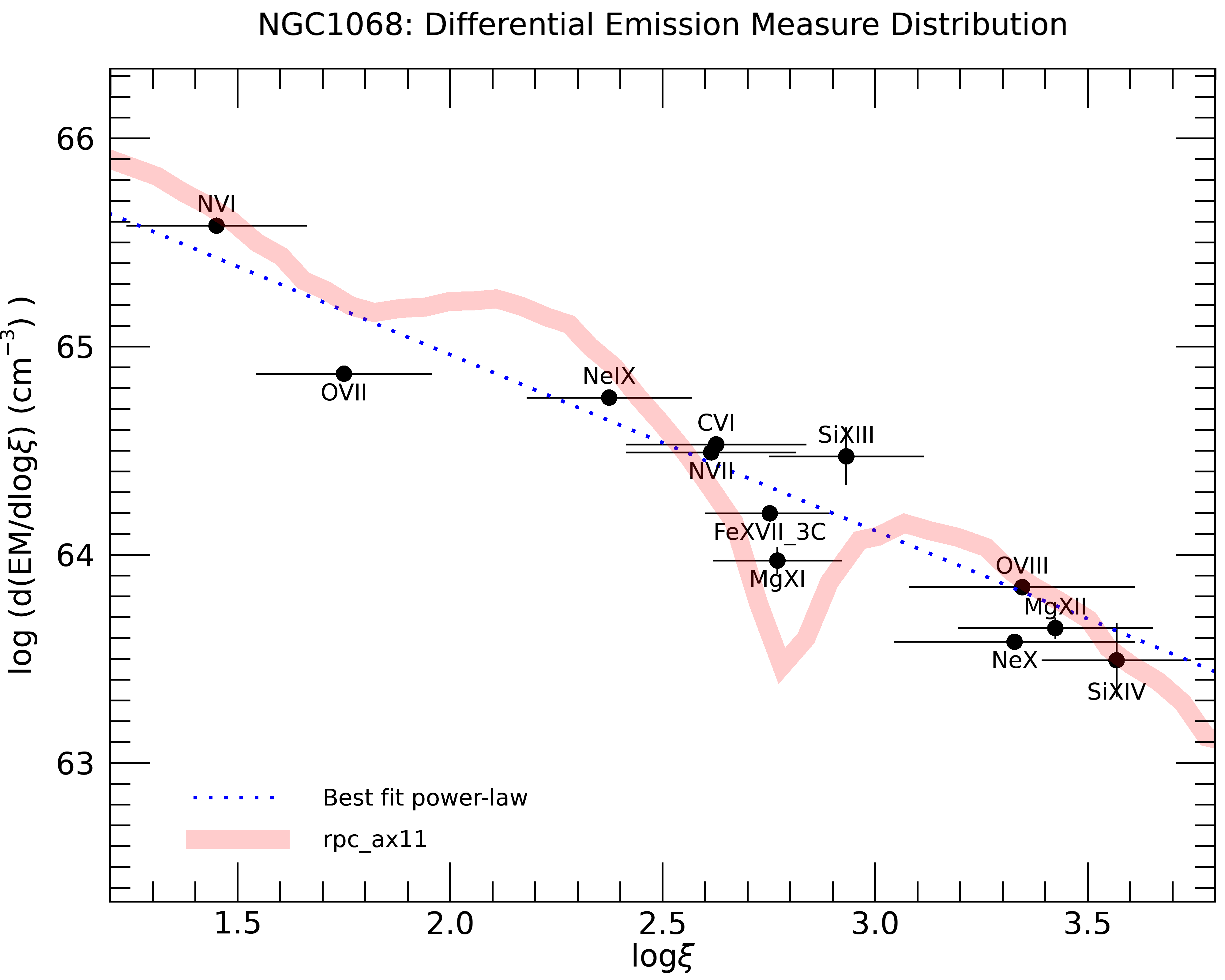}
\includegraphics[width=0.48\textwidth]{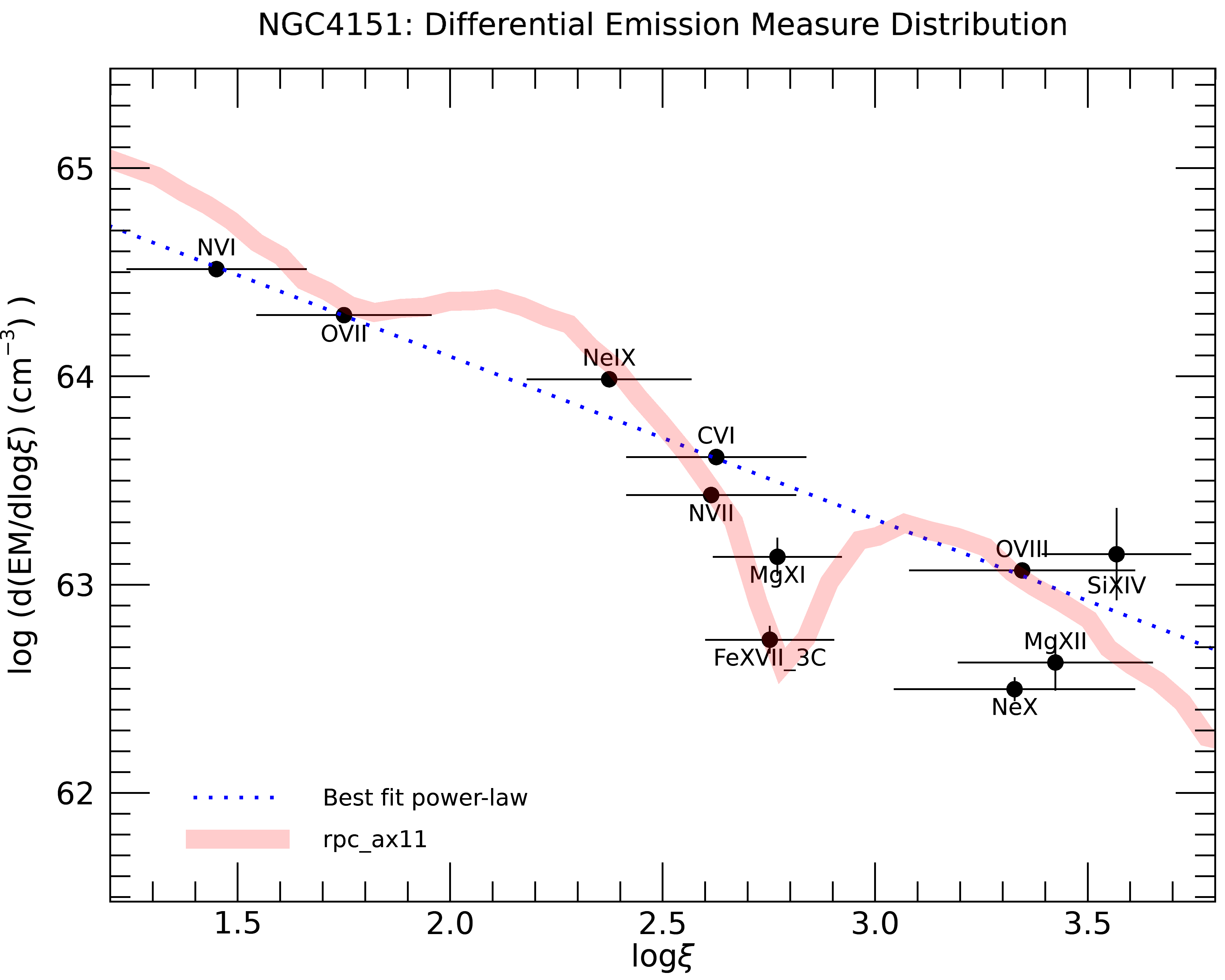}
\caption{\label{EMD1} Observed DEM distributions (black circles) for NGC~1068 (\textit{left}) and NGC~4151 (\textit{right}). The best fit power-law is over-imposed, along with the DEM distribution predicted for a RPC gas which best reproduces the data (red shaded area): see Table~\ref{tab2}.}
\end{figure*}

Our analysis detects at least 3 different emission lines (out of the 12 selected to build the DEM as in Table~\ref{tab1}) in 26 sources. Only in one source, NGC~1068, all the 12 lines are detected: the derived DEM distribution is shown in Fig.~\ref{EMD1} (left panel). The observed DEM evidently appears as a power-law distribution: a linear regression gives a slope of $\sim-0.85$ (the blue dotted line in the plot). In the same figure, the RPC prediction is over-imposed, the one that best reproduce the data, chosen among those derived for different illuminating SEDs and metallicities (see Fig.~\ref{EMD_RPC}). The correspondence between the observed DEM and the distribution predicted for a RPC gas is impressive.

\begin{table*}

  \caption{The CHRESOS sources with at least 3 emission lines detected out of the 12 listed in Table~\ref{tab1}. }
  \label{tab2}
  \begin{tabular}{l|c|c|c|c|c|c} \hline \hline
Source & $\log L_{2-10}$ & Lines & Best fit & DEM ($\log \xi/\mathrm{erg\,cm\, s^{-1}}=2$) & DEM Slope  & $\Omega/4\pi$\\
(1) & (2) & (3) & (4) & (5) & (6) & (7)\\
\hline
NGC~1068 & 42.93 & 12 & \texttt{ax11} & $65.086\pm0.004$ & $-0.846\pm0.006$ & $1.567\pm0.014$\\
\rule{0pt}{3ex}NGC~4151 & 42.31 & 11 & \texttt{ax11} & $64.284\pm0.005$ & $-0.782\pm0.006$ & $1.030\pm0.012$\\
\rule{0pt}{3ex}NGC~1365 & 42.32 & 8 & \texttt{ax11} & $64.162\pm0.019$ & $-0.53\pm0.04$ & $0.76\pm0.03$\\
\rule{0pt}{3ex}NGC~5548 & 43.14 & 7 & \texttt{ax11} & $64.77\pm0.03$ & $-0.87\pm0.05$ & $0.47\pm0.03$\\
Circinus & 42.63 & 7 & \texttt{ax07} & $63.52\pm0.08$ & $-0.2\pm0.2$ & $0.085^{+0.017}_{-0.014}$\\
NGC~7582 & 43.48 & 7 & \texttt{ax11} & $63.86\pm0.05$ & $-0.45\pm0.09$ & $0.026\pm0.003$\\
\rule{0pt}{3ex}ESO362-G018 & 42.96 & 6 & \texttt{ax11} & $64.57\pm0.04$ & $-0.71\pm0.06$ & $0.45\pm0.04$\\
MRK~3 & 43.67 & 6 & \texttt{ax11} & $64.99\pm0.05$ & $-0.58\pm0.07$ & $0.23^{+0.03}_{-0.02}$\\
NGC~4507 & 43.51 & 6 & \texttt{ax11} & $64.61\pm0.05$ & $-0.84\pm0.08$ & $0.138^{+0.017}_{-0.015}$\\
NGC~5506 & 42.99 & 6 & \texttt{ax11}  & $64.08\pm0.04$ & $-0.49\pm0.09$ & $0.135^{+0.013}_{-0.012}$\\
\rule{0pt}{3ex}IRAS05189-2524 & 43.40 & 5 & \texttt{ax11} & $65.47\pm0.10$ & $-0.6\pm0.2$ & $1.3\pm0.3$\\
NGC~424 & 43.77 & 5 & \texttt{aion16} & $64.16\pm0.06$ & $-0.91\pm0.09$ & $0.027^{+0.004}_{-0.003}$\\
\rule{0pt}{3ex}ESO138-G01 & 44.09 & 4 & \texttt{ax11} & $64.50\pm0.08$ & $-0.65\pm0.12$ & $0.028^{+0.006}_{-0.005}$\\
MRK~477 & 43.26 & 4 & \texttt{ax11} & $65.1\pm0.2$ & $-0.9\pm0.3$ & $0.8^{+0.4}_{-0.3}$\\
NGC~777 & -- & 4 & \texttt{aion19} & $66.05\pm0.04$ & $-1.84\pm0.13$ & --\\
NGC~1052 & 41.62 & 4 & \texttt{ax11} & $63.47\pm0.11$ & $-0.43\pm0.19$ & $0.77^{+0.22}_{-0.17}$\\
NGC~5643 & 42.43 & 4 & \texttt{ax11} & $63.04\pm0.14$ & $-0.72\pm0.18$ & $0.045^{+0.017}_{-0.012}$\\
NGC~6240 & 44.75 & 4 & \texttt{aion19} & $66.03\pm0.06$ & $-2.1\pm0.2$ & $0.21\pm0.03$\\
\rule{0pt}{3ex}H0557-385 & 44.08 & 3 & \texttt{ax11} & $65.8\pm0.8$ & $0.30\pm0.13$ & $0.6^{+3.0}_{-0.5}$\\
IRAS13197-1627 & 43.41 & 3 & \texttt{aion13} & $64.82\pm0.11$ & $-0.63\pm0.18$ & $0.28^{+0.08}_{-0.06}$\\
MRK~231 & 42.59 & 3 & \texttt{ax07} & $65.46\pm0.19$ & $-0.2\pm0.6$ & $8^{+4}_{-3}$\\
MRK~704 & 43.33 & 3 & \texttt{ax11} & $65.20\pm0.10$ & $-0.8\pm0.3$ & $0.81^{+0.21}_{-0.17}$\\
NGC~1320 & 42.85 & 3 & \texttt{aion19} & $64.09\pm0.17$ & $-0.4\pm0.4$ & $0.19^{+0.09}_{-0.06}$\\
NGC~3393 & 42.63 & 3 & \texttt{aion13} & $64.60\pm0.17$ & $-0.7\pm0.2$ & $1.0^{+0.5}_{-0.3}$\\
NGC~4388 & 43.05 & 3 & \texttt{metals3}  & $64.13\pm0.14$ & $-0.68\pm0.18$ & $0.13^{+0.05}_{-0.04}$\\
UGC~1214 & -- & 3 & \texttt{metals3} & $64.80\pm0.16$ & $-0.7\pm0.2$ & --\\
\hline \hline
\end{tabular}

\medskip
\begin{minipage}{\textwidth}
Notes. (1) Name of the source. (2) Logarithm of the unabsorbed intrinsic 2-10 keV luminosity (cgs). All values taken from \citet{Ricci2017} except for Mrk~231 \citep{Teng2014}, NGC~1320 \citep{Balokovic2014}. (3) Number of emission lines detected out of the 12 listed in Table~\ref{tab1}. (4) RPC DEM that best fit the observed DEM (\texttt{ax\#\#}: SED variant with the reported X-ray slope - \texttt{aion\#\#}: SED variant with the reported $\alpha_{ion}$ - \texttt{metals\#}: metal overabundance by the reported factor. See text for details on the adopted SEDs and metallicities). (5) Normalization of the observed DEM (cgs) at $\log \xi/\mathrm{erg\,cm\, s^{-1}}=2$, assuming a linear fit with slope fixed to -1.19 (see text for details). (6) Best-fit slope for a power-law fit of the derived DEM. (7) Covering factor of the emitting gas as derived from (2) and (5), see text for details.
\end{minipage}
\end{table*}

\begin{figure*}
\includegraphics[width=0.48\textwidth]{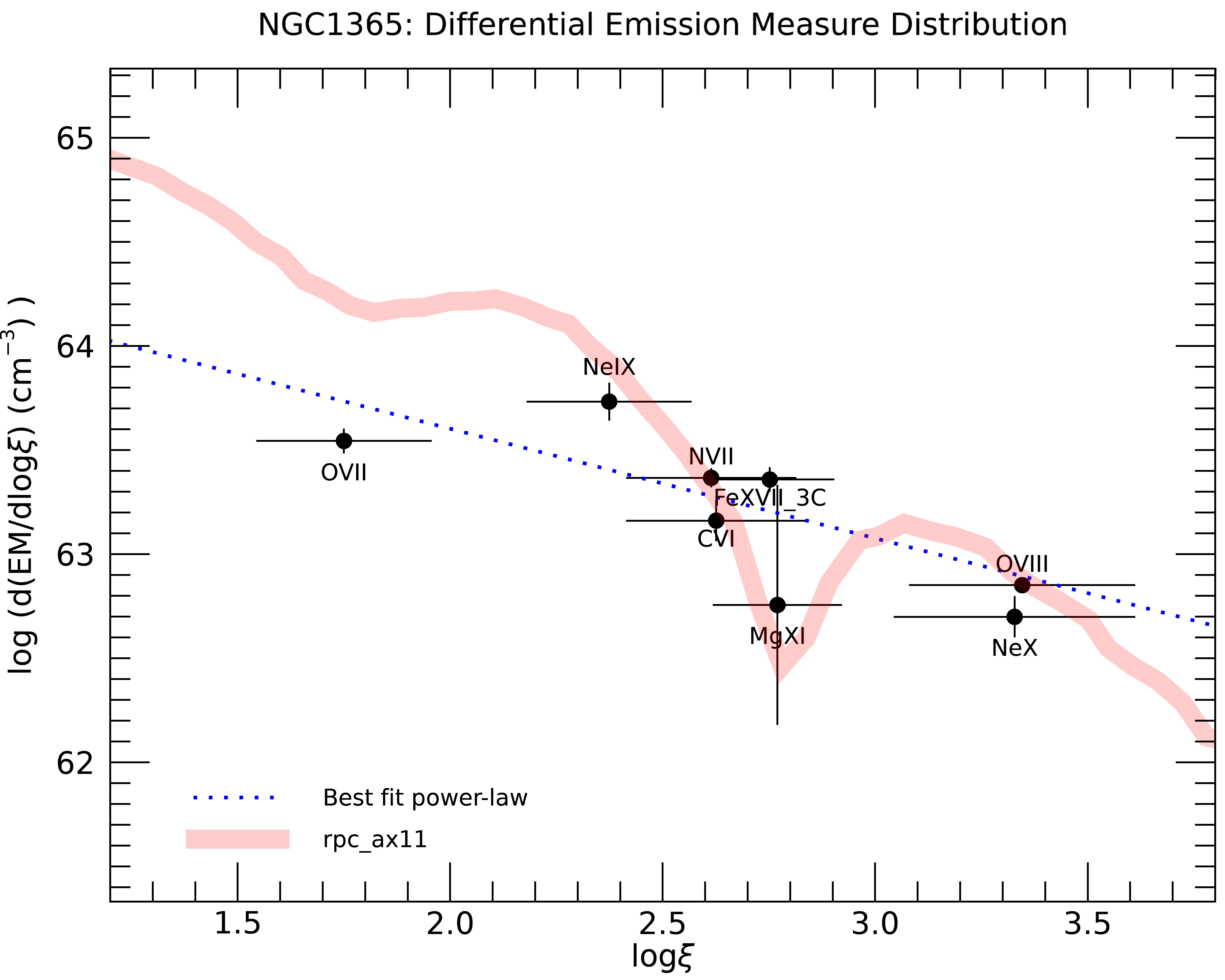}
\includegraphics[width=0.48\textwidth]{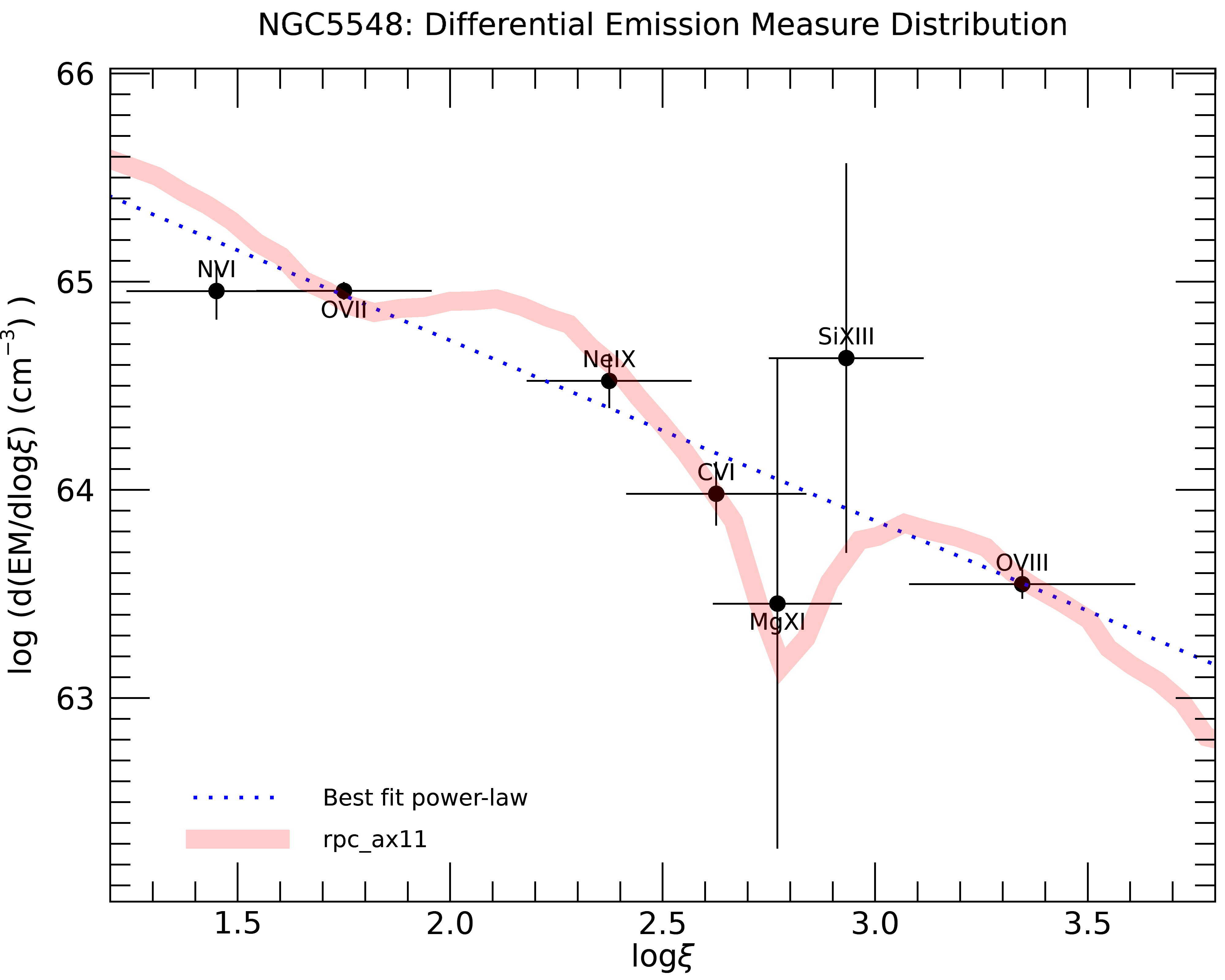}
\includegraphics[width=0.48\textwidth]{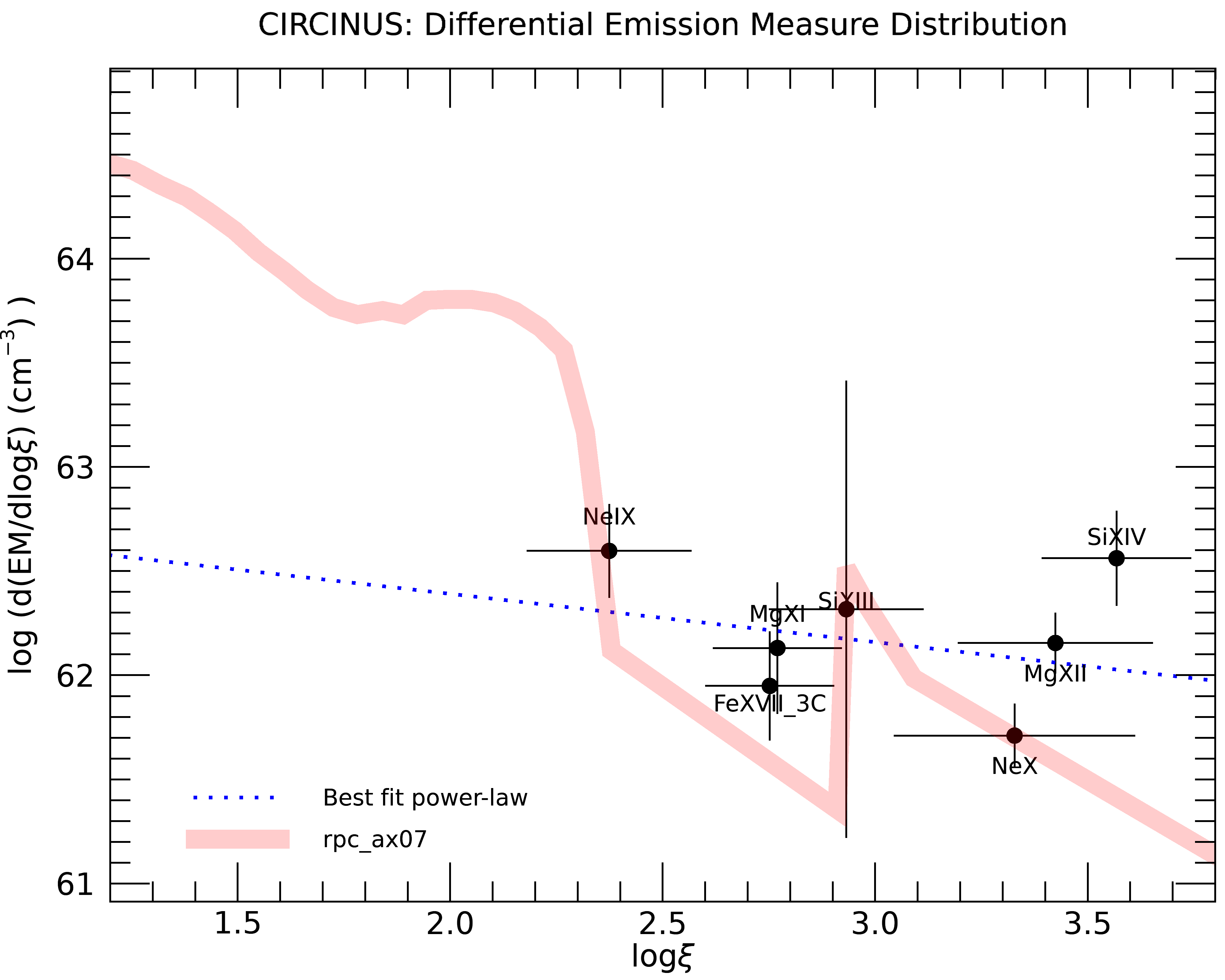}
\includegraphics[width=0.48\textwidth]{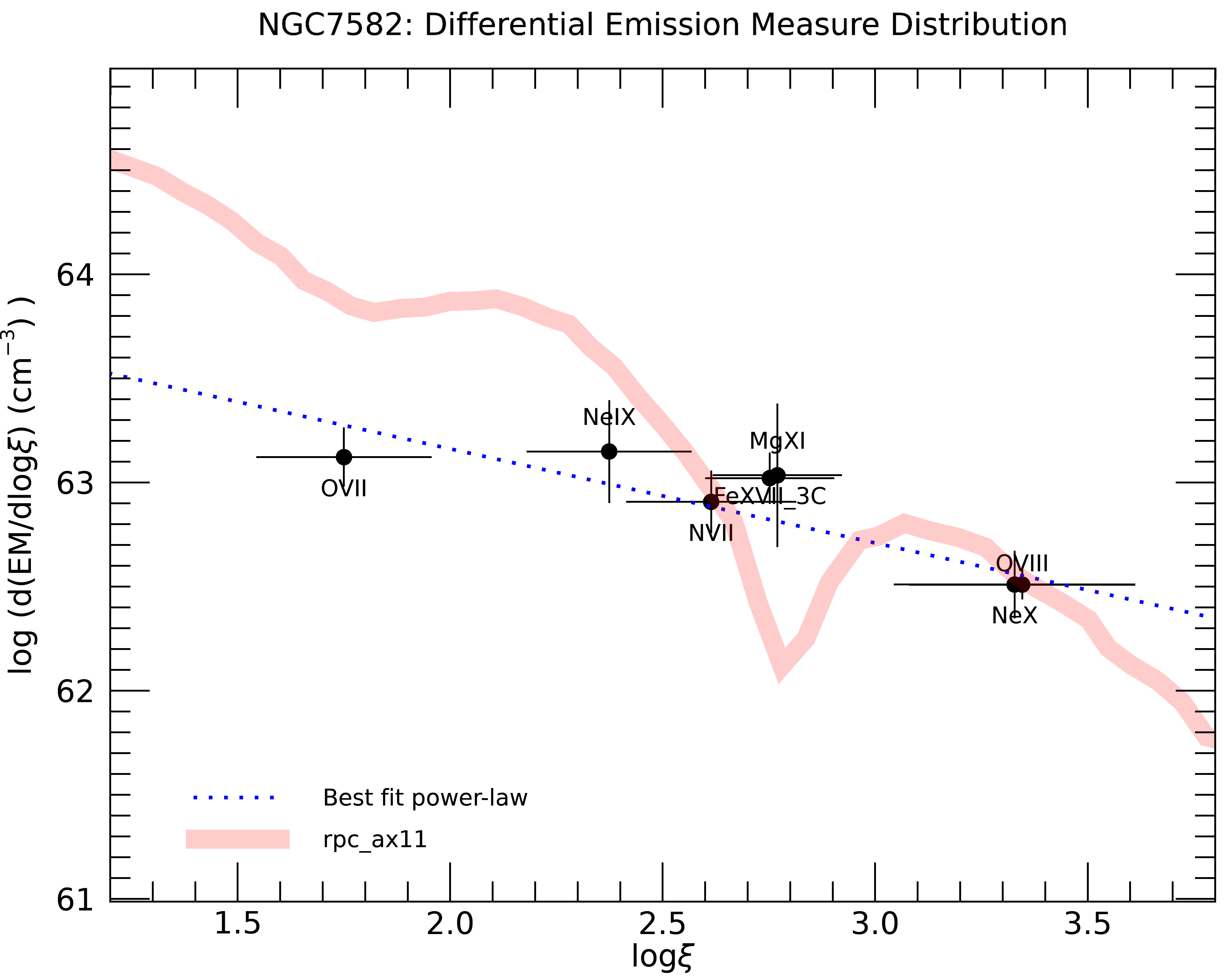}
\caption{\label{EMD2}Same as Fig.~\ref{EMD1}, but for the sources with 7-8 detected lines: see Table~\ref{tab2}.}
\end{figure*}

\begin{figure*}
\includegraphics[width=0.32\textwidth]{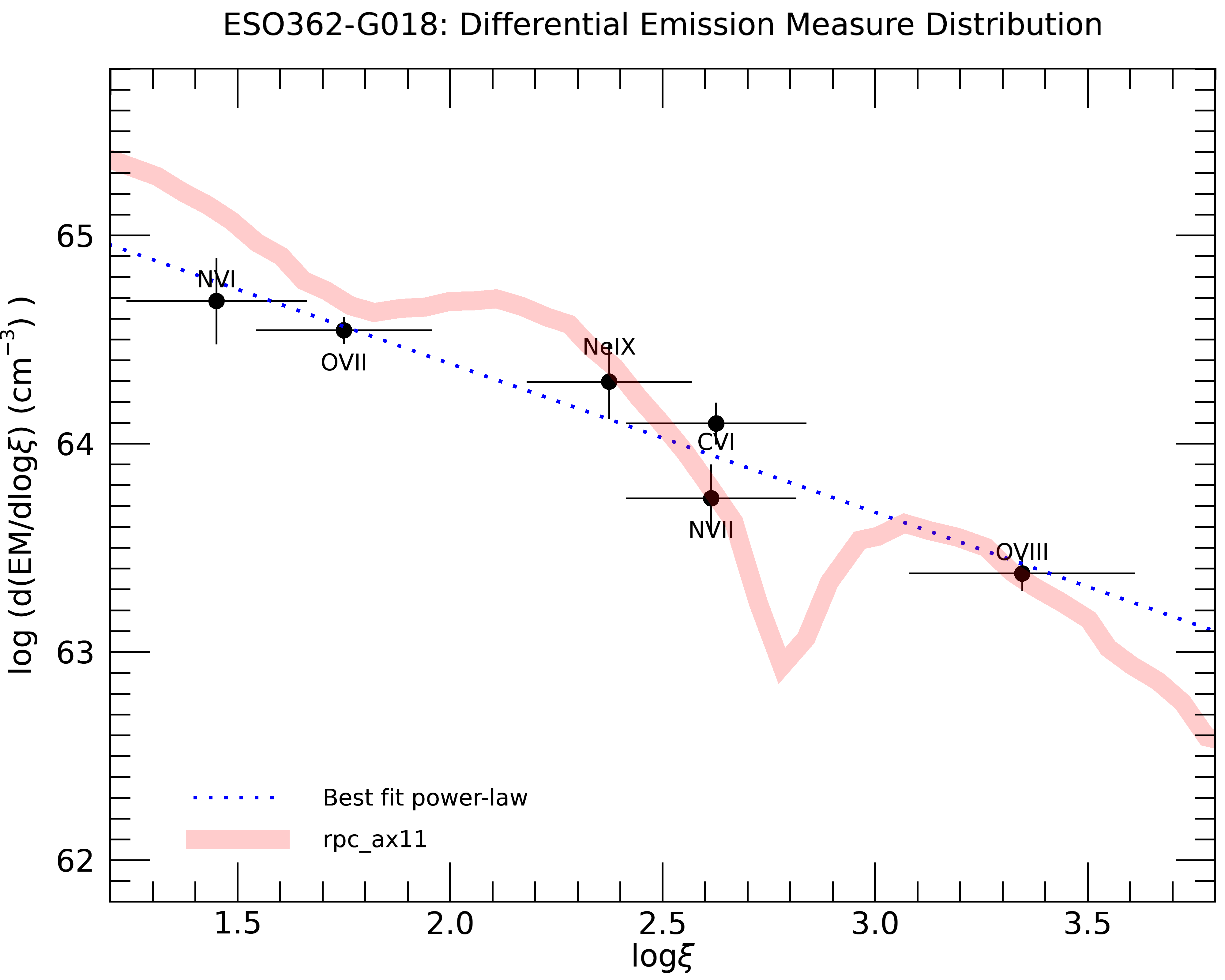}
\includegraphics[width=0.32\textwidth]{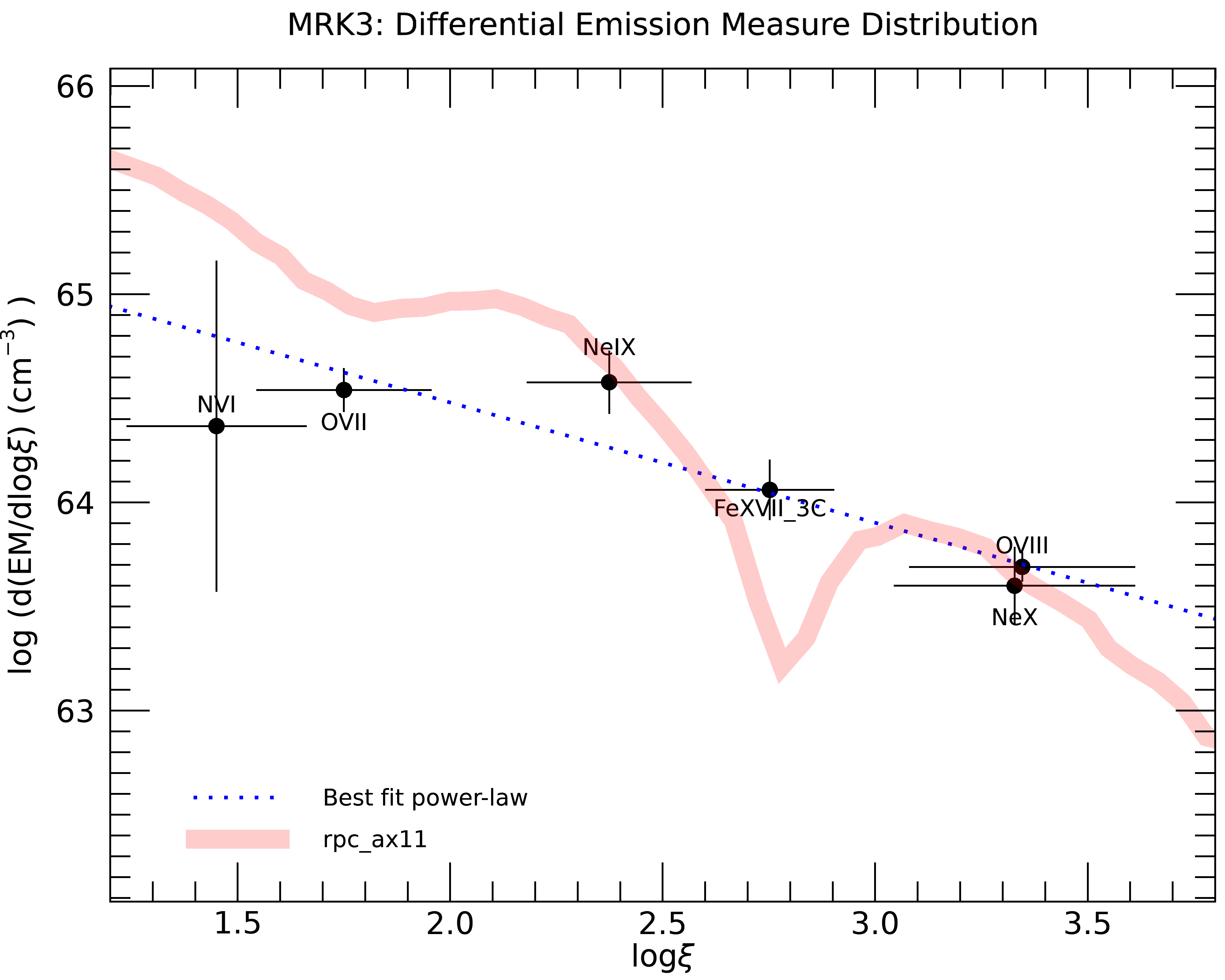}
\includegraphics[width=0.32\textwidth]{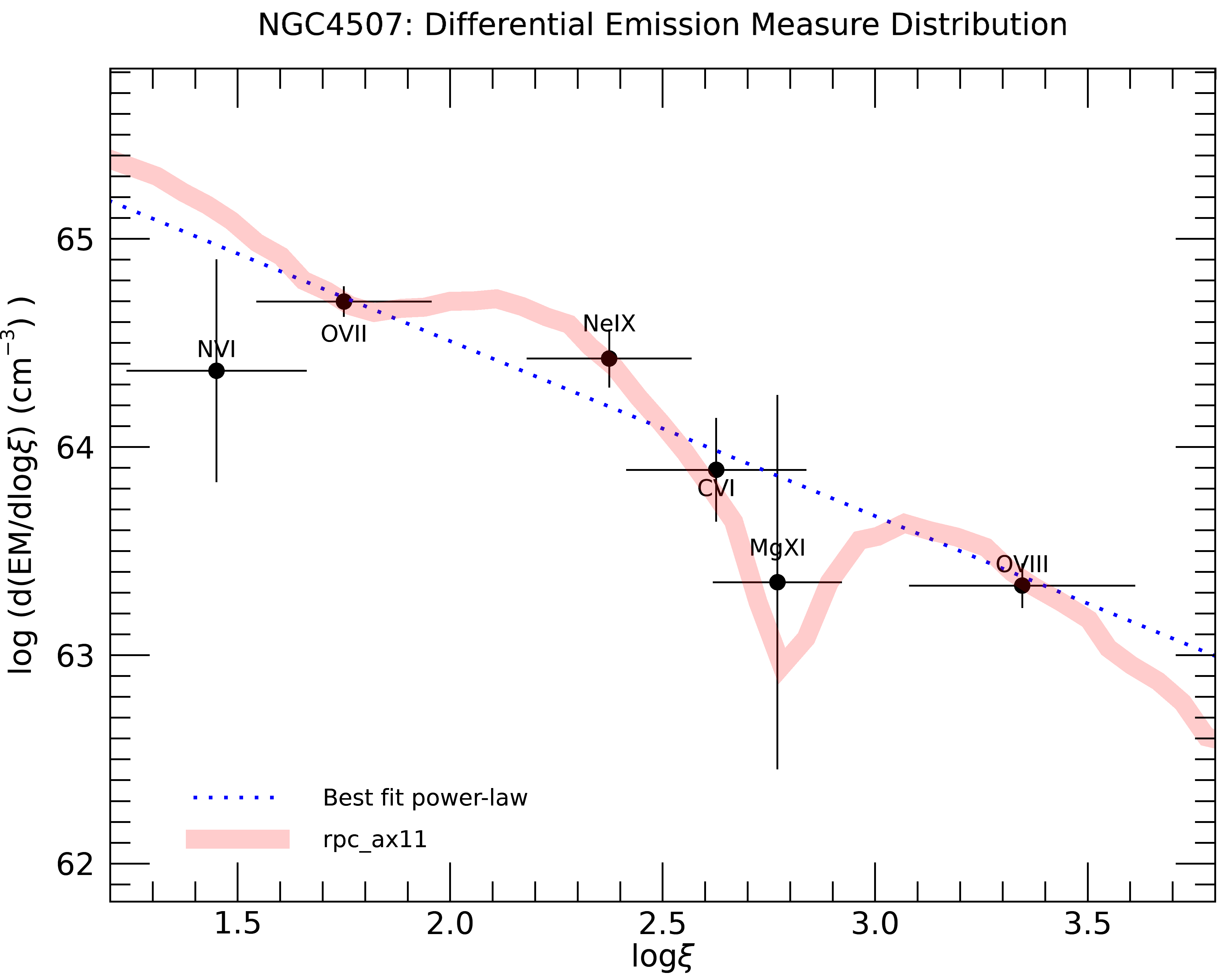}
\includegraphics[width=0.32\textwidth]{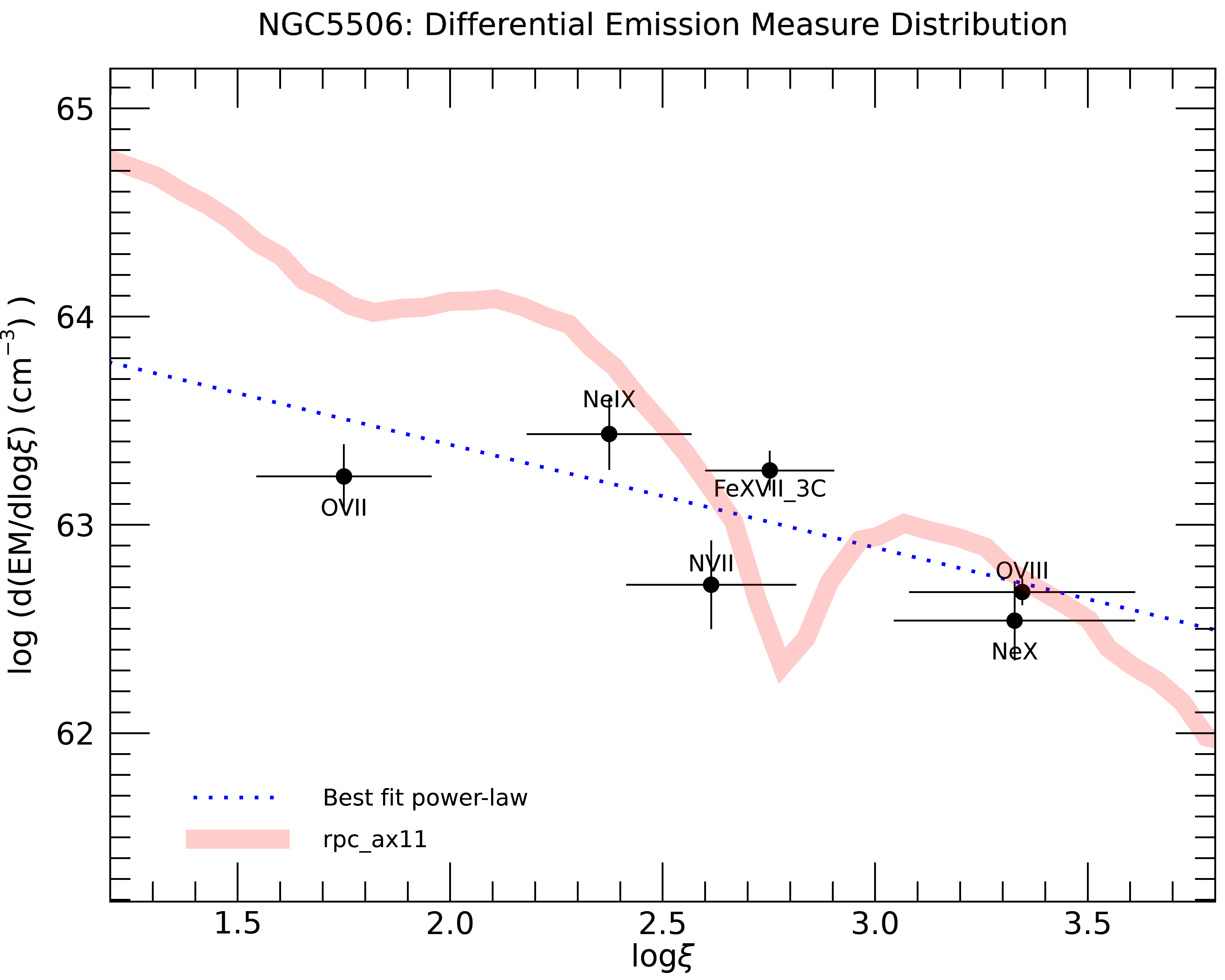}
\includegraphics[width=0.32\textwidth]{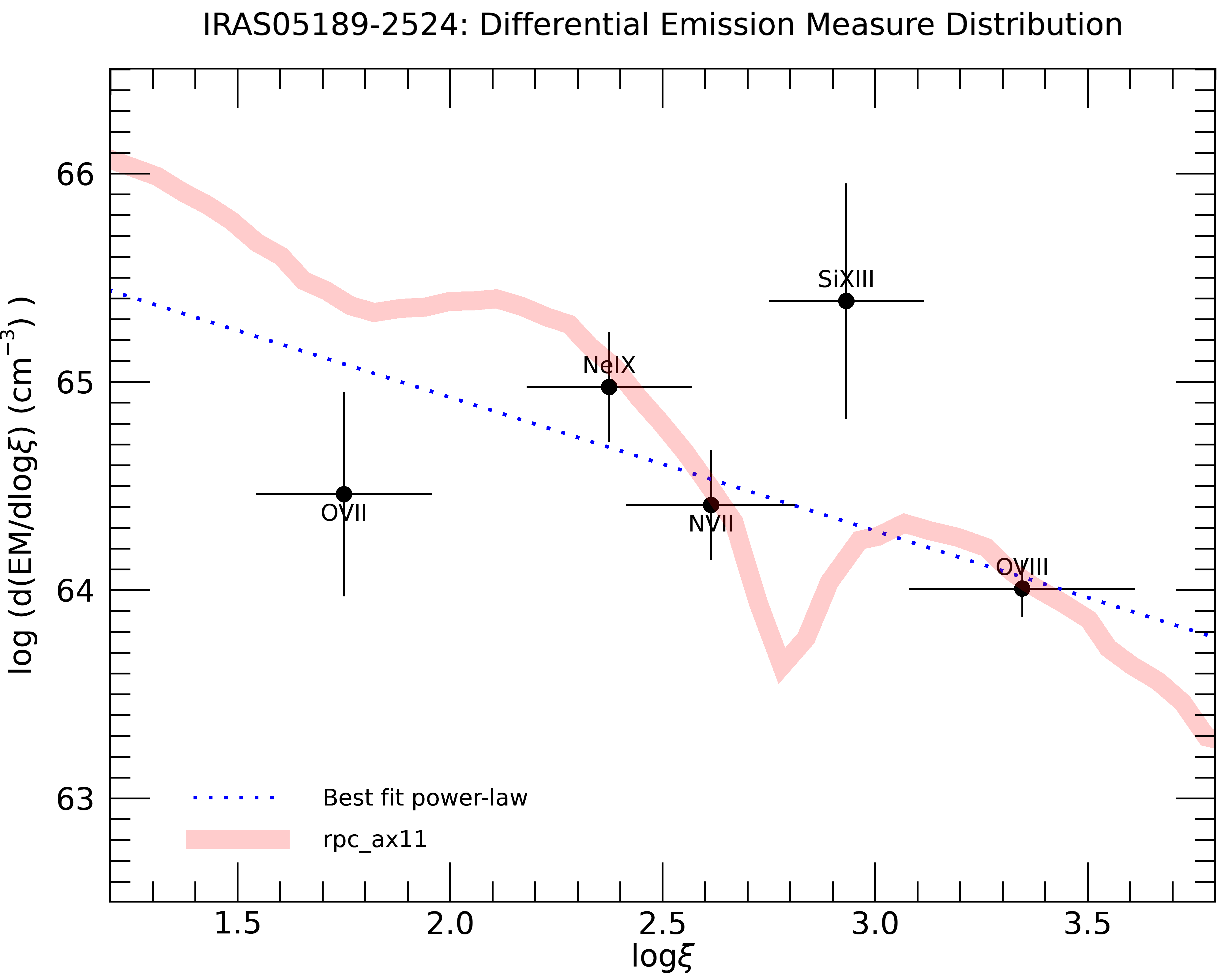}
\includegraphics[width=0.32\textwidth]{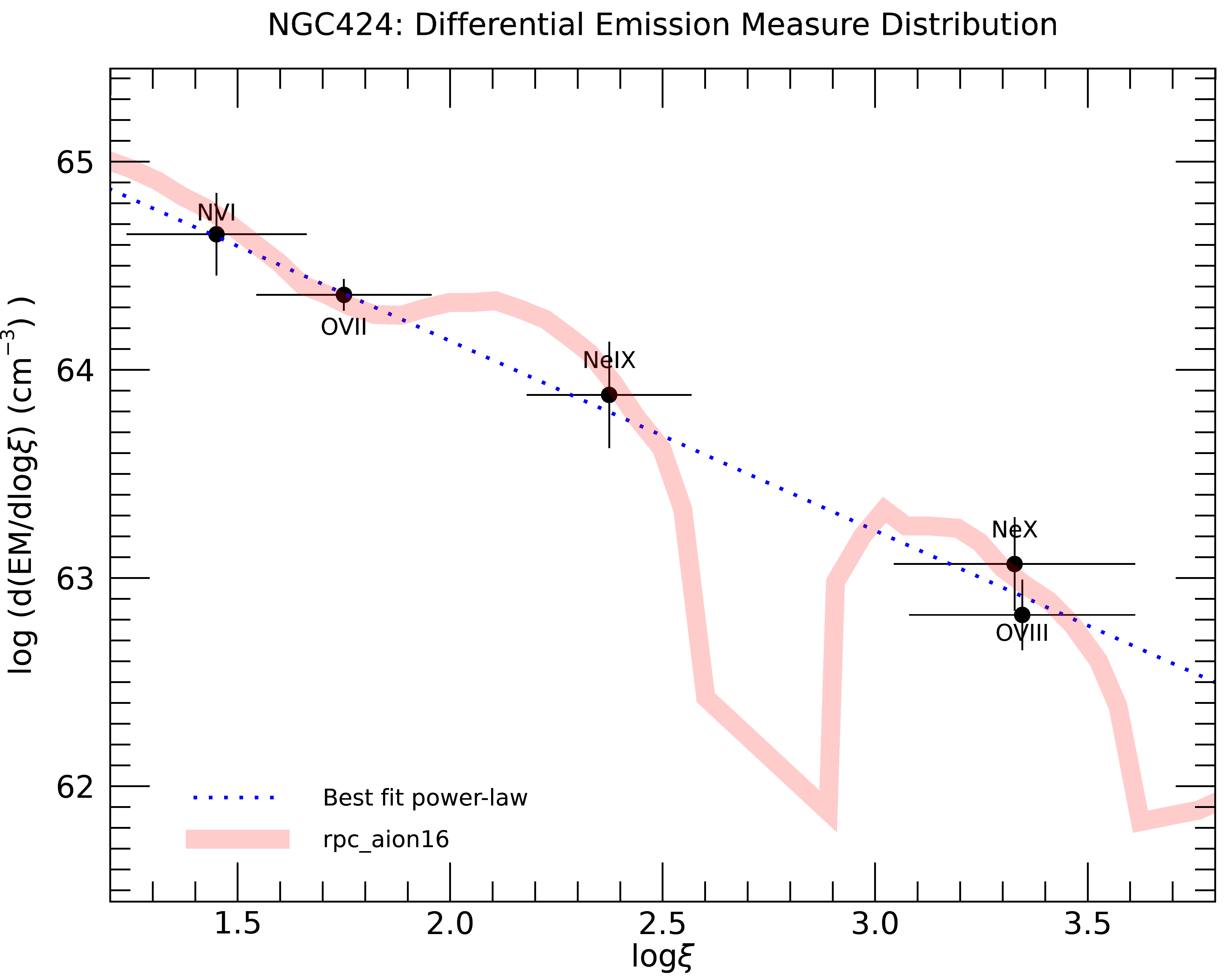}
\caption{\label{EMD3}Same as Fig.~\ref{EMD1}, but for the sources with 5-6 detected lines: see Table~\ref{tab2}.}
\end{figure*}

\begin{figure*}
\includegraphics[width=0.32\textwidth]{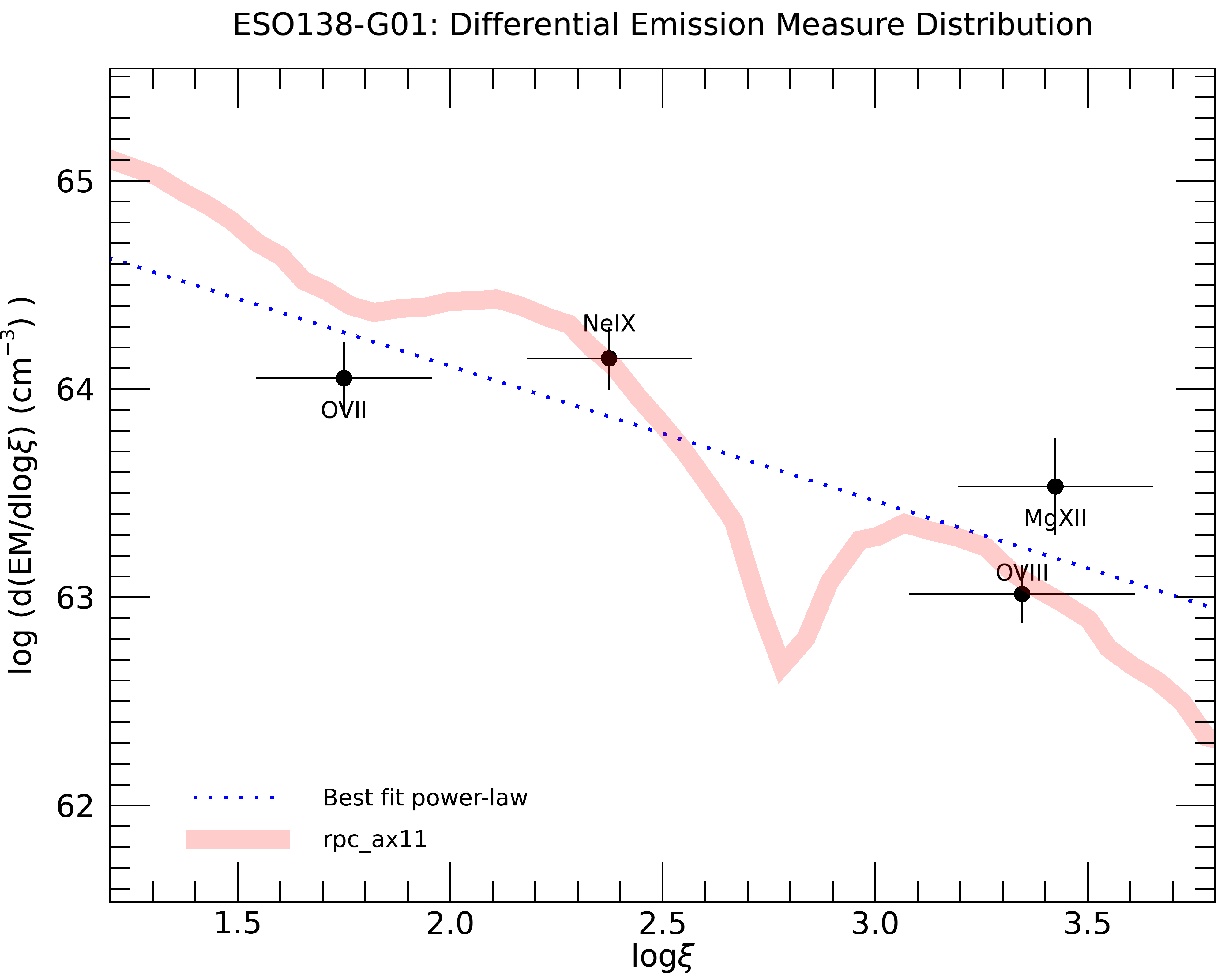}
\includegraphics[width=0.32\textwidth]{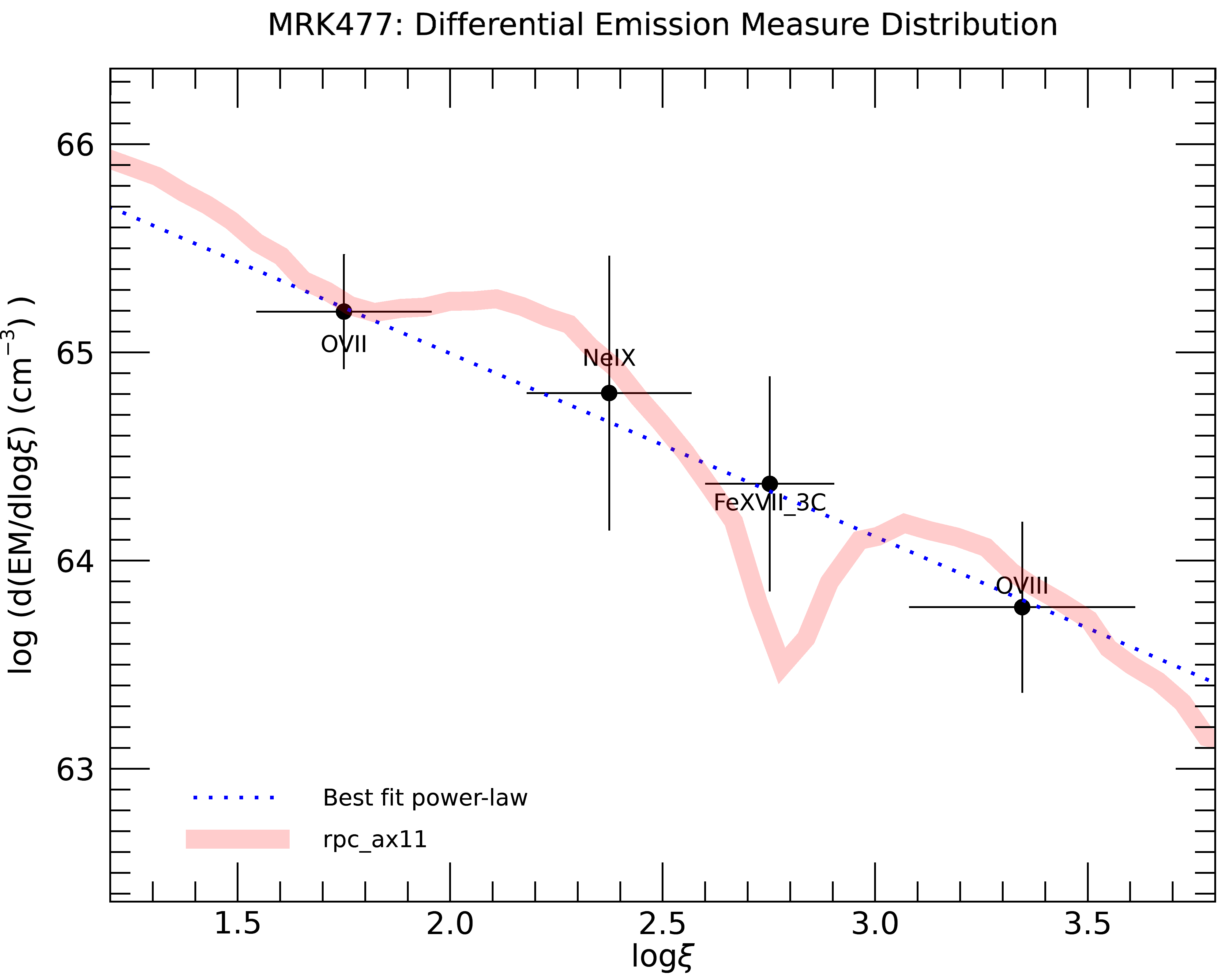}
\includegraphics[width=0.32\textwidth]{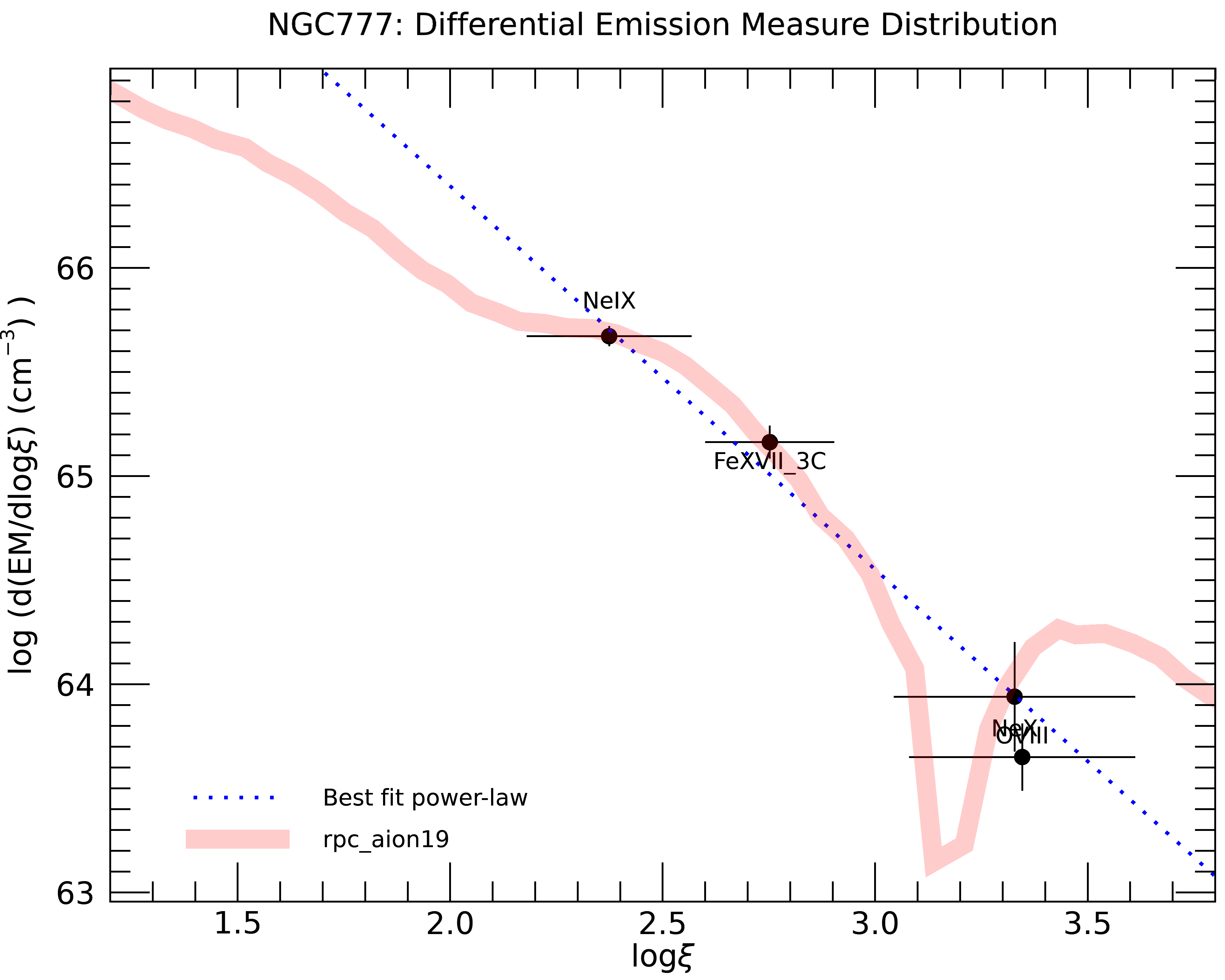}
\includegraphics[width=0.32\textwidth]{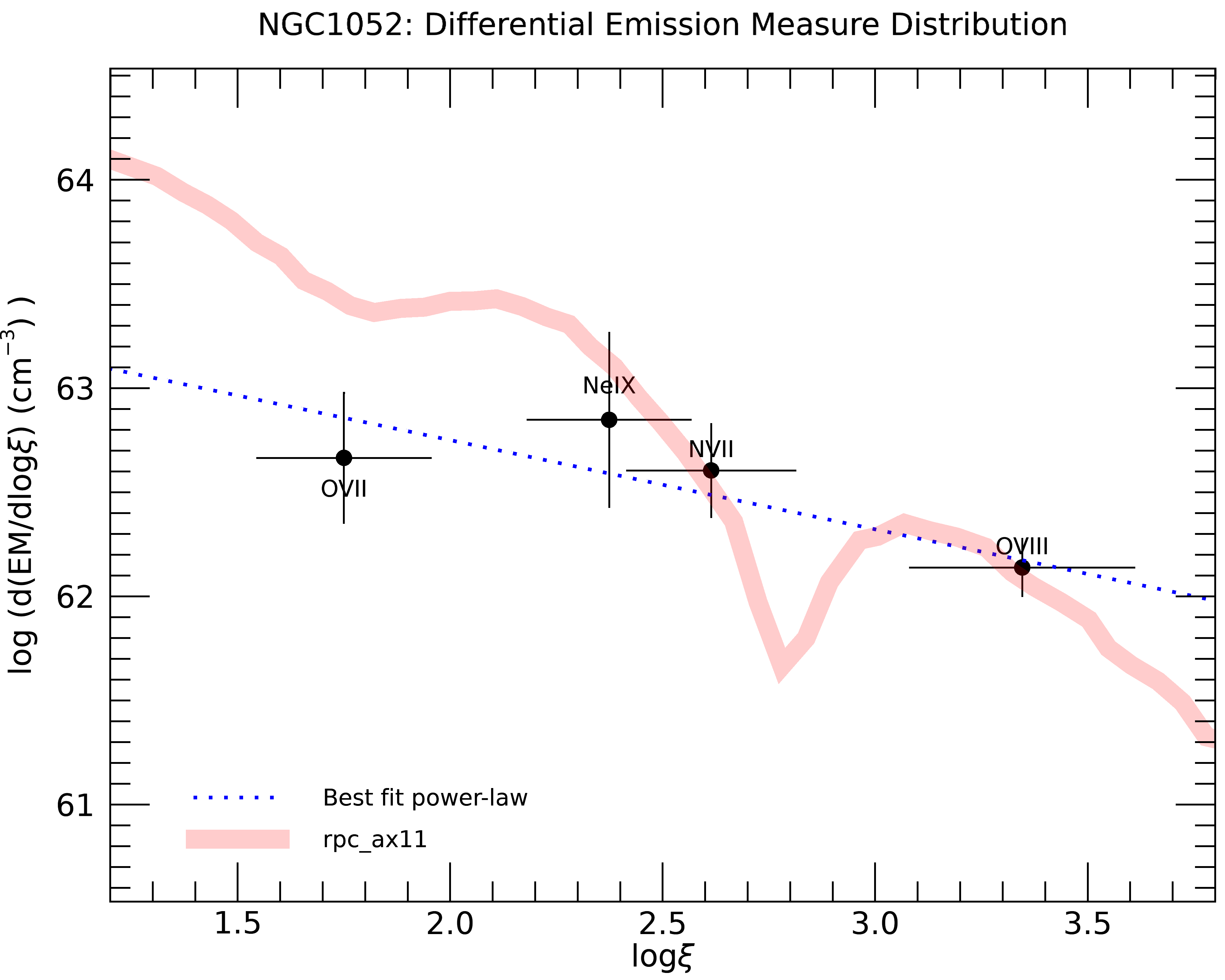}
\includegraphics[width=0.32\textwidth]{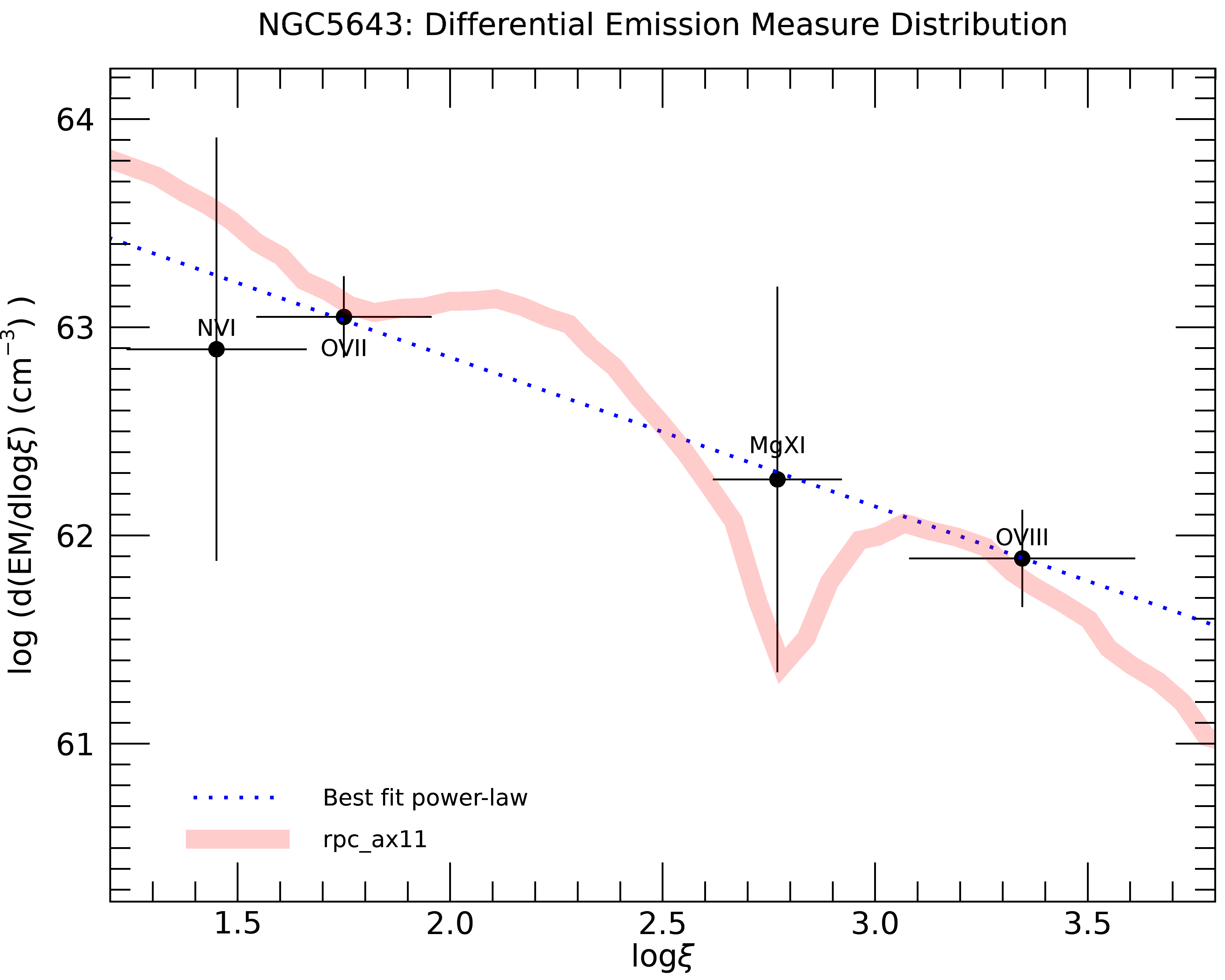}
\includegraphics[width=0.32\textwidth]{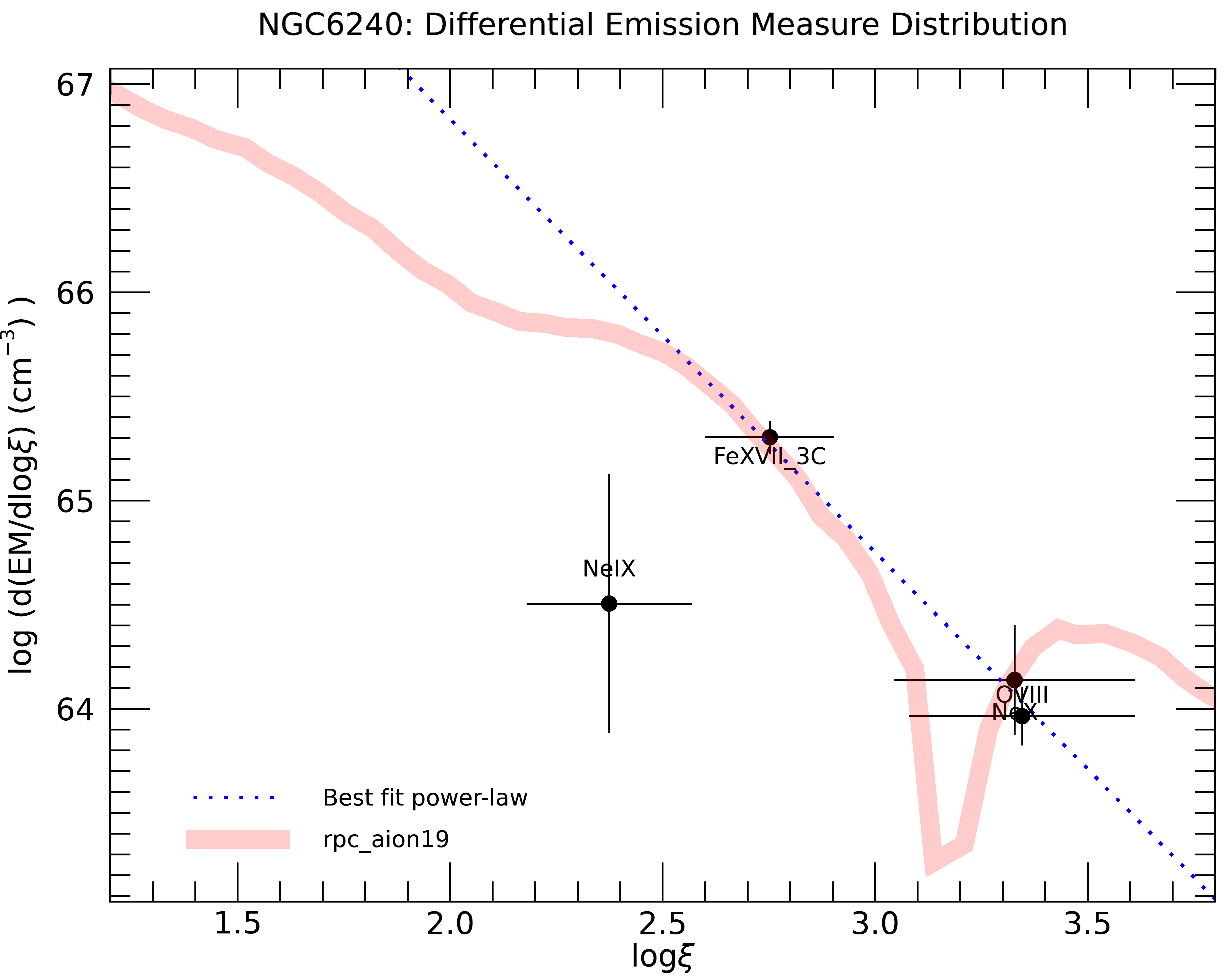}
\caption{\label{EMD4}Same as Fig.~\ref{EMD1}, but for the sources with 4 detected lines: see Table~\ref{tab2}.}
\end{figure*}

The observed DEM distribution of NGC~4151 is almost as good as that of NGC~1068, with 11 detected lines (see right panel of Fig.~\ref{EMD1}). Their overall shapes are very similar, and again in extremely good agreement with the RPC predictions. The DEM distributions of the other 24 sources are shown in Fig.~\ref{EMD2}-\ref{EMD3}-\ref{EMD4}, grouped on the basis of the number of detected lines. Most of the DEM distributions are in good agreement with the RPC predictions, with few notable examples of flatter observed curves. Circinus is indeed a source which shows an extremely flat DEM, confirming earlier results by \citet{Sako2000}: however, this source is characterized by an exceptionally high Galactic column density, which severely limits the S/N of the low-energy spectrum. On the other hand, note that none of the observed DEM distributions are instead steeper than RPC. The only exception is NGC~777, but the best fit slope seems driven by the lack of observed low ionization lines: all the detected emission lines are indeed in perfect agreement with the RPC prediction (see Fig.~\ref{EMD4}).

\begin{figure*}
\includegraphics[width=0.32\textwidth]{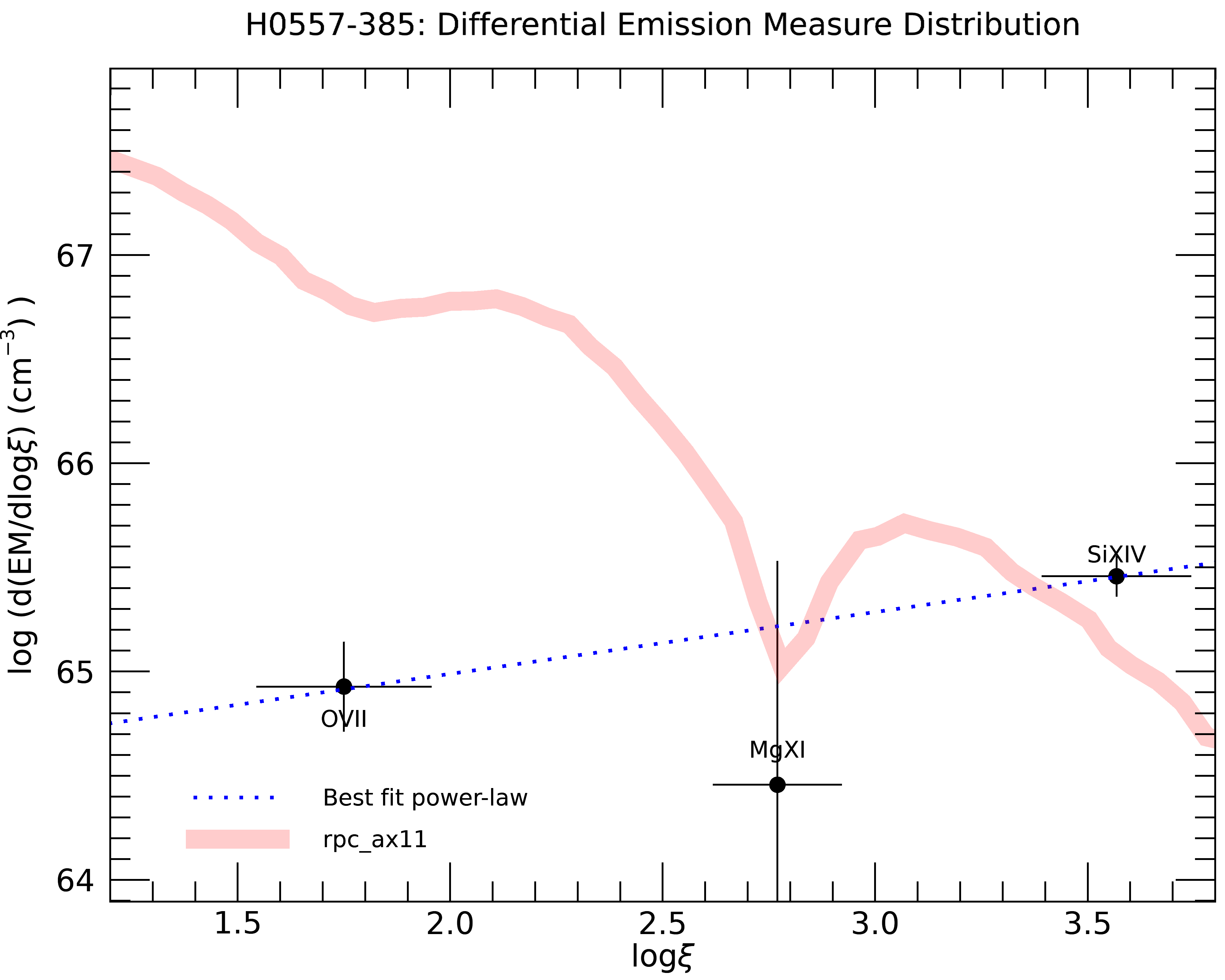}
\includegraphics[width=0.32\textwidth]{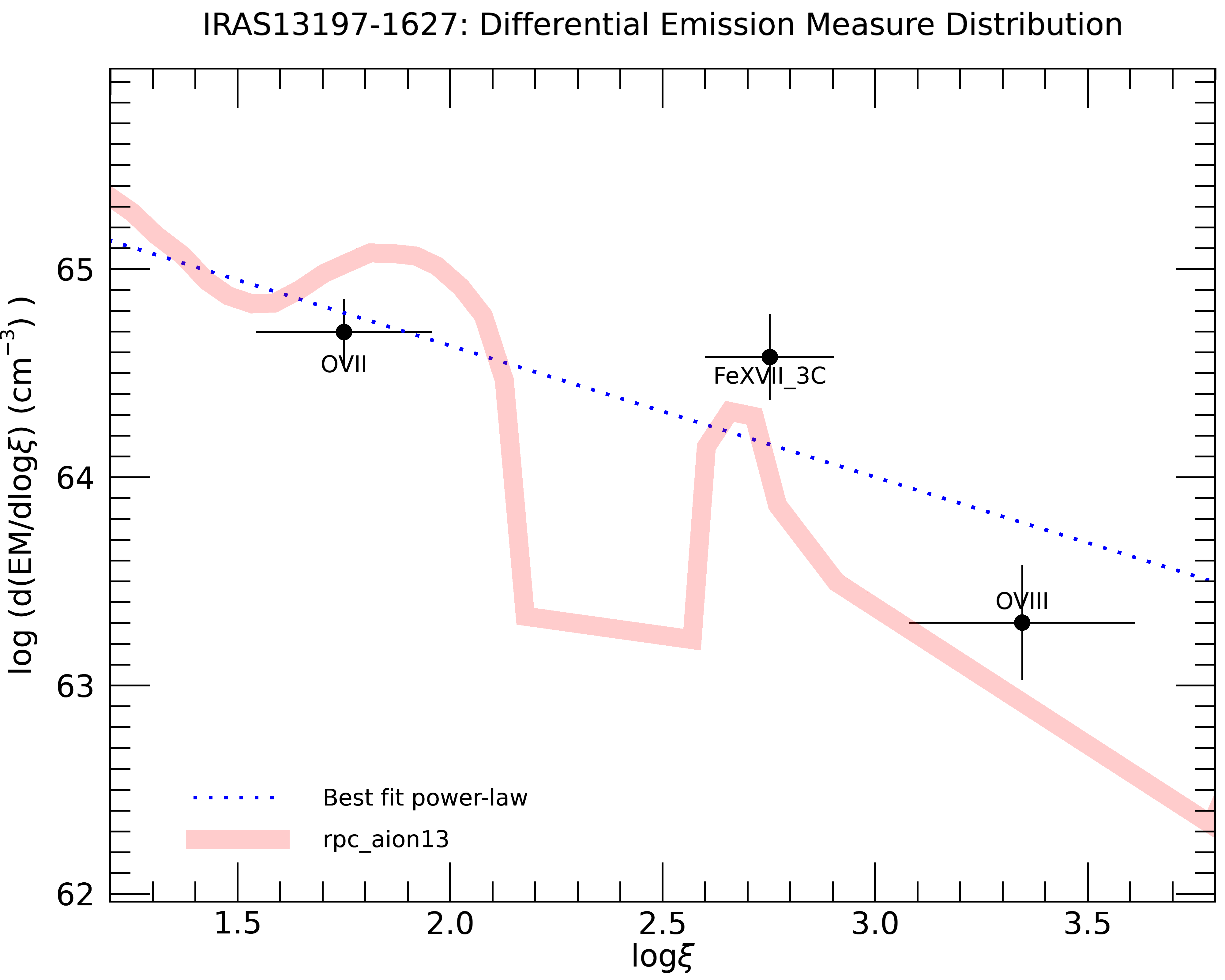}
\includegraphics[width=0.32\textwidth]{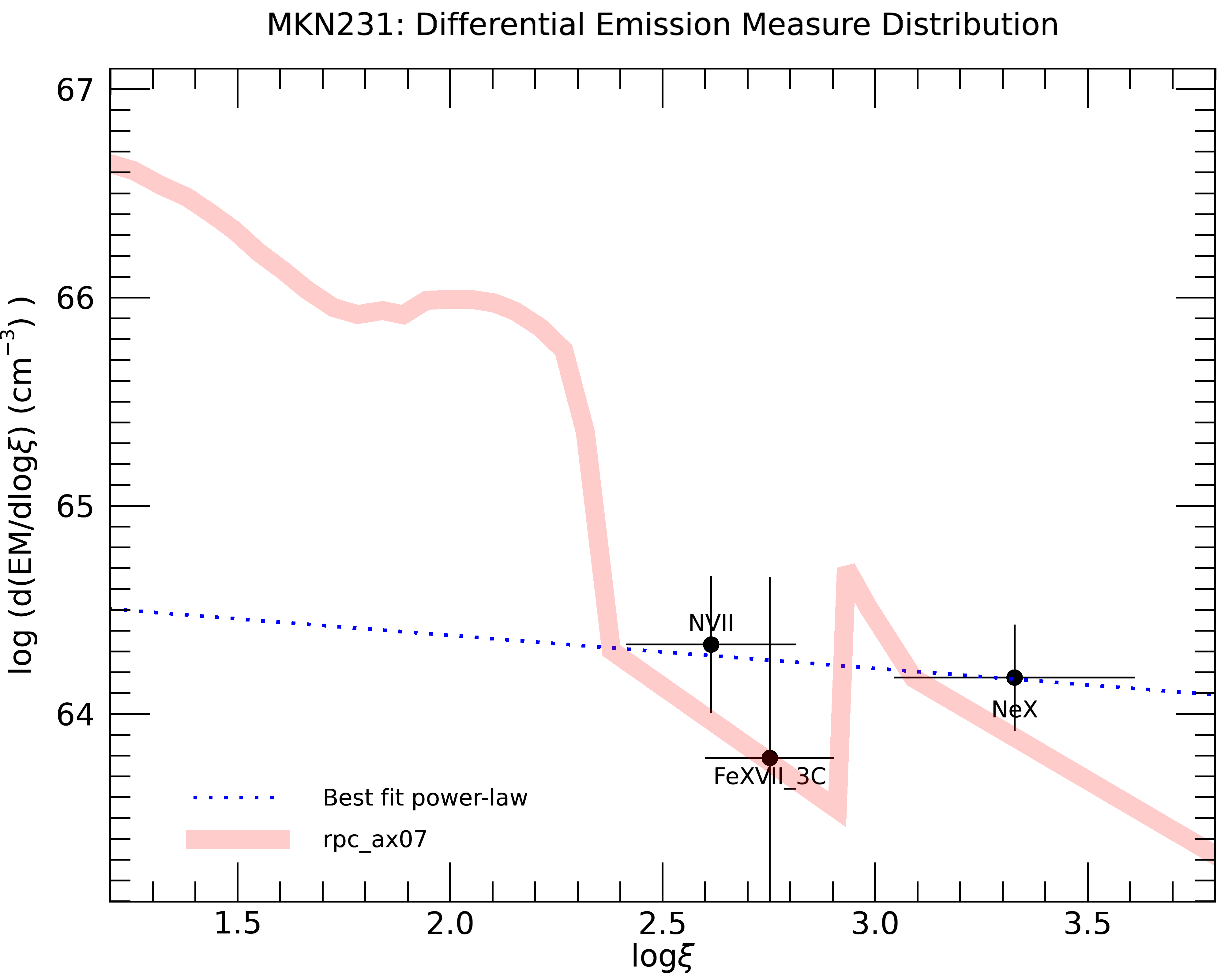}
\includegraphics[width=0.32\textwidth]{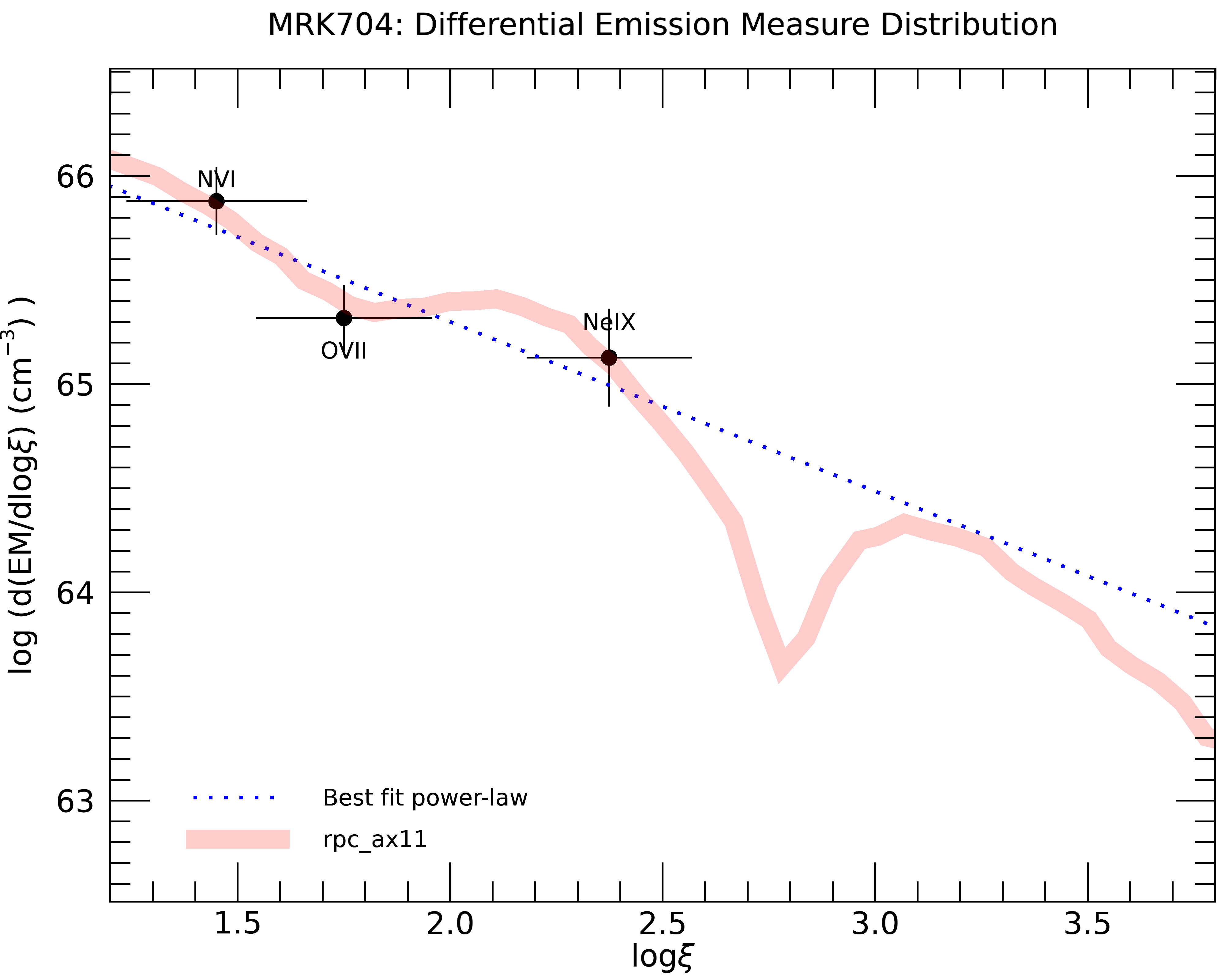}
\includegraphics[width=0.32\textwidth]{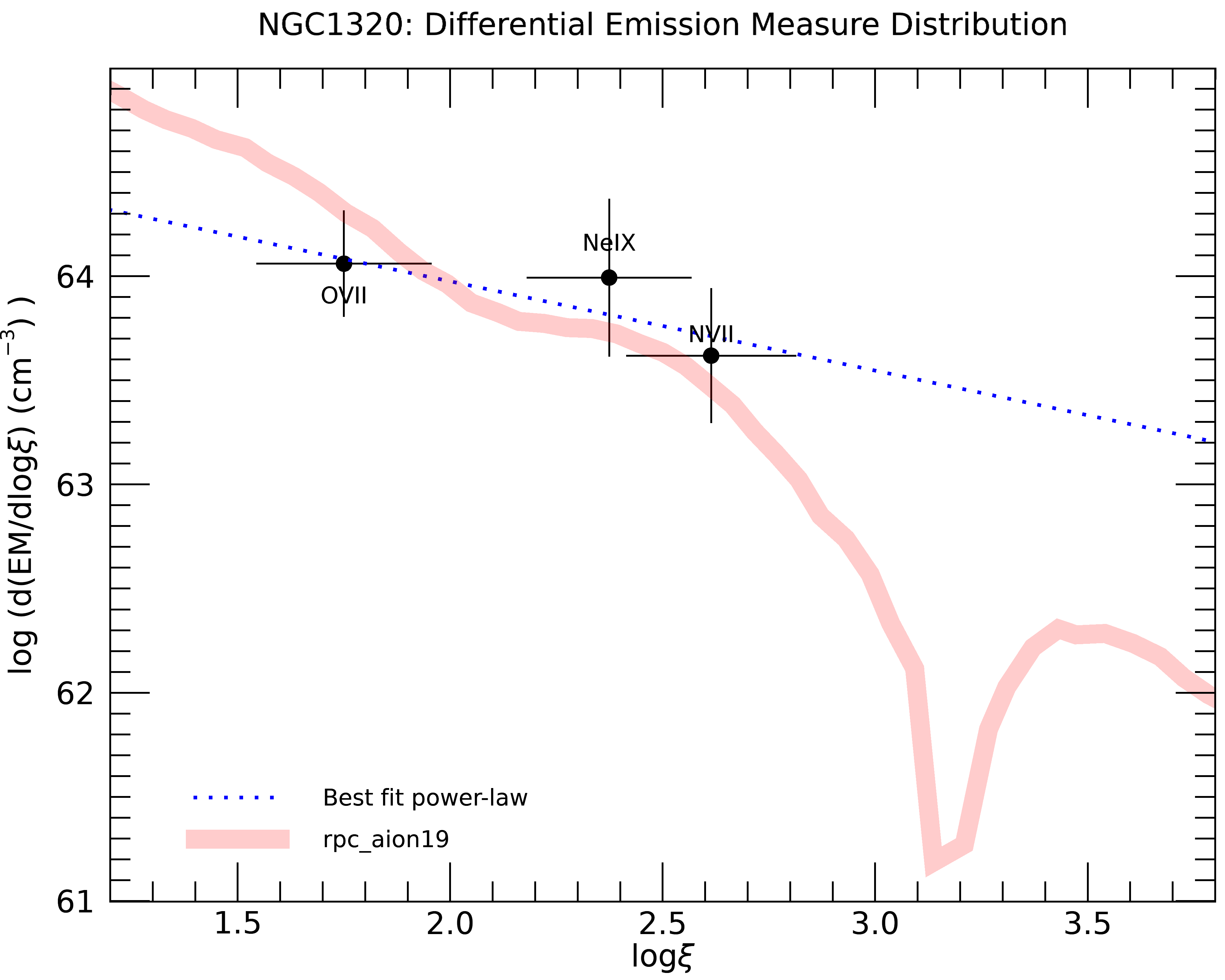}
\includegraphics[width=0.32\textwidth]{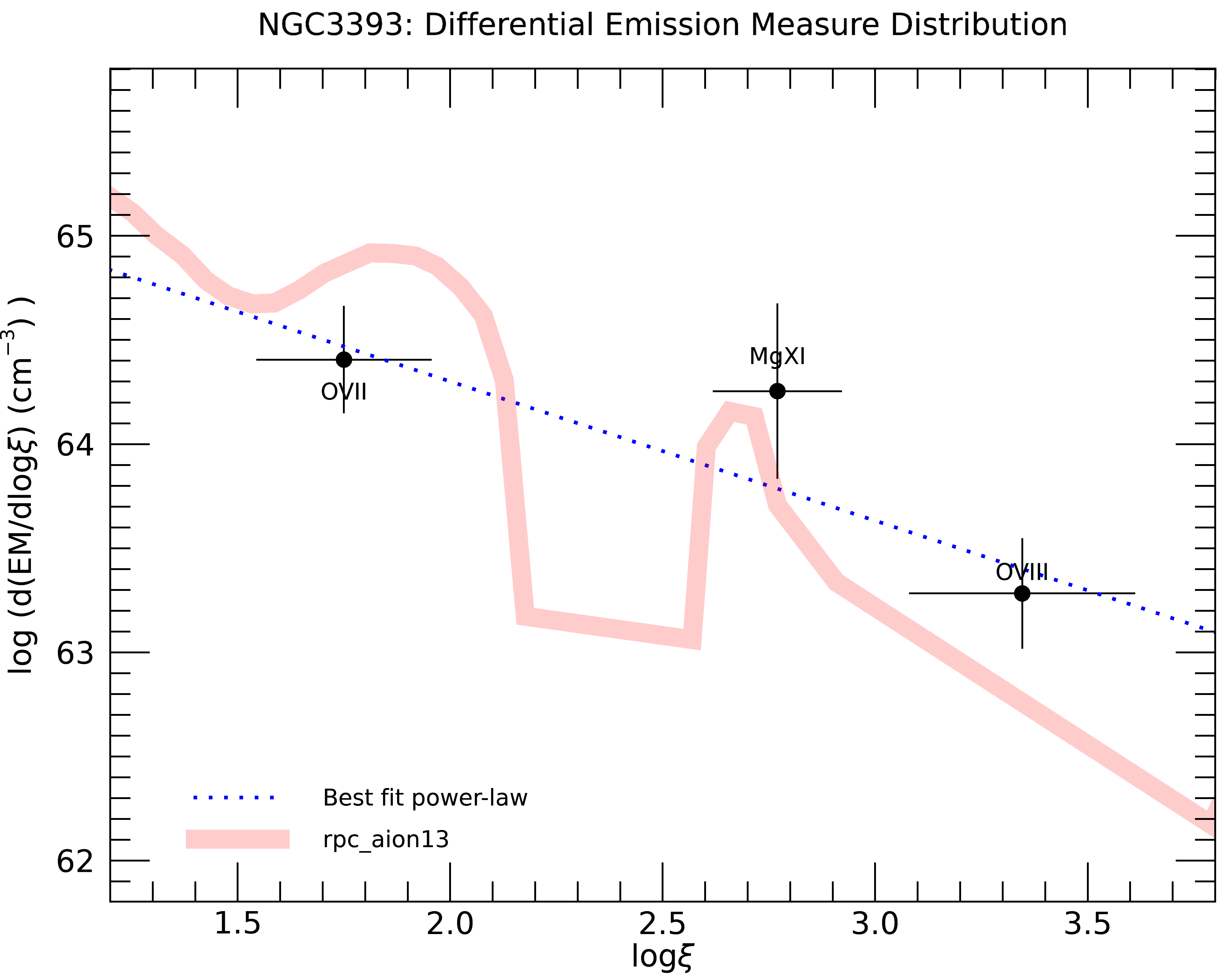}
\includegraphics[width=0.32\textwidth]{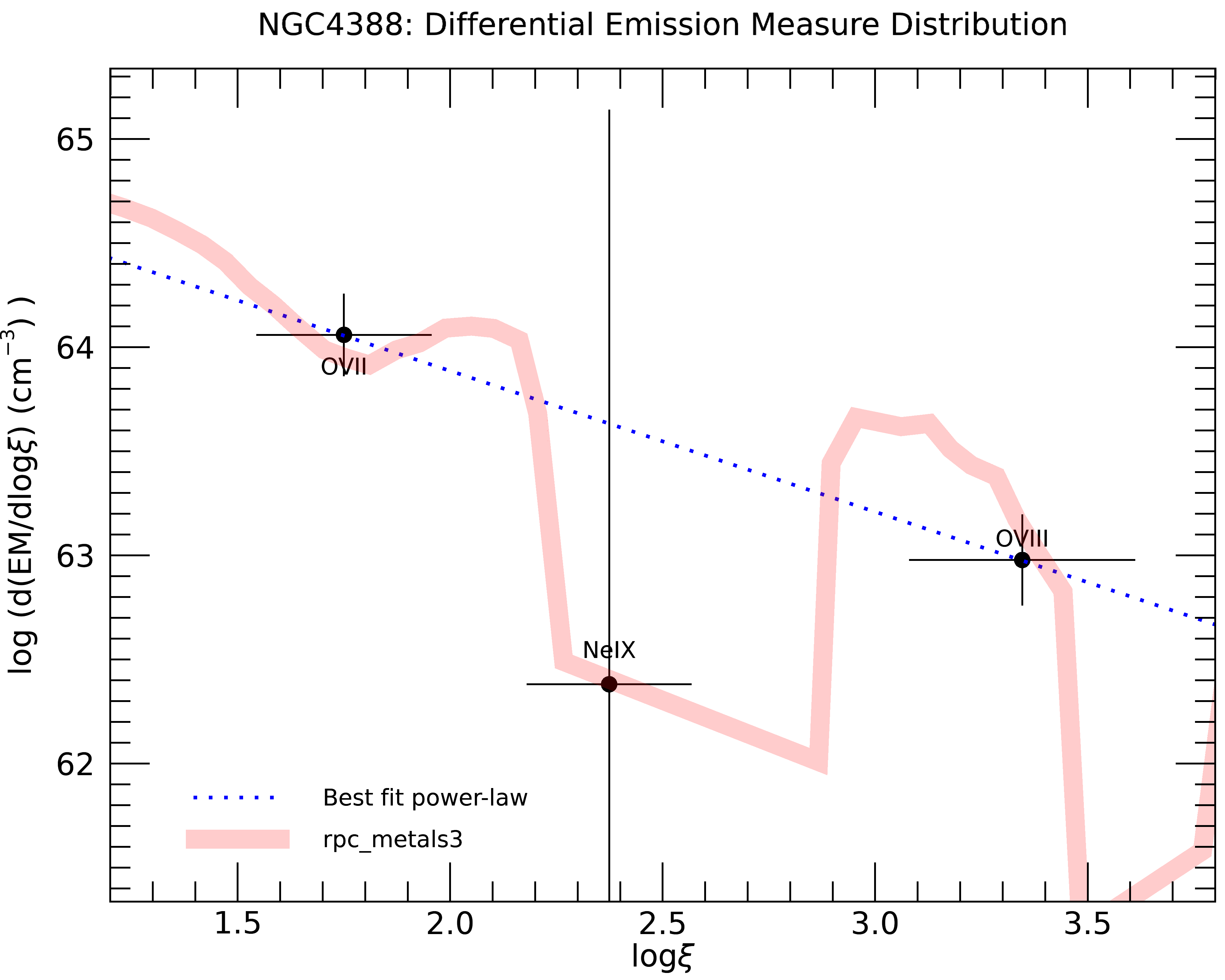}
\includegraphics[width=0.32\textwidth]{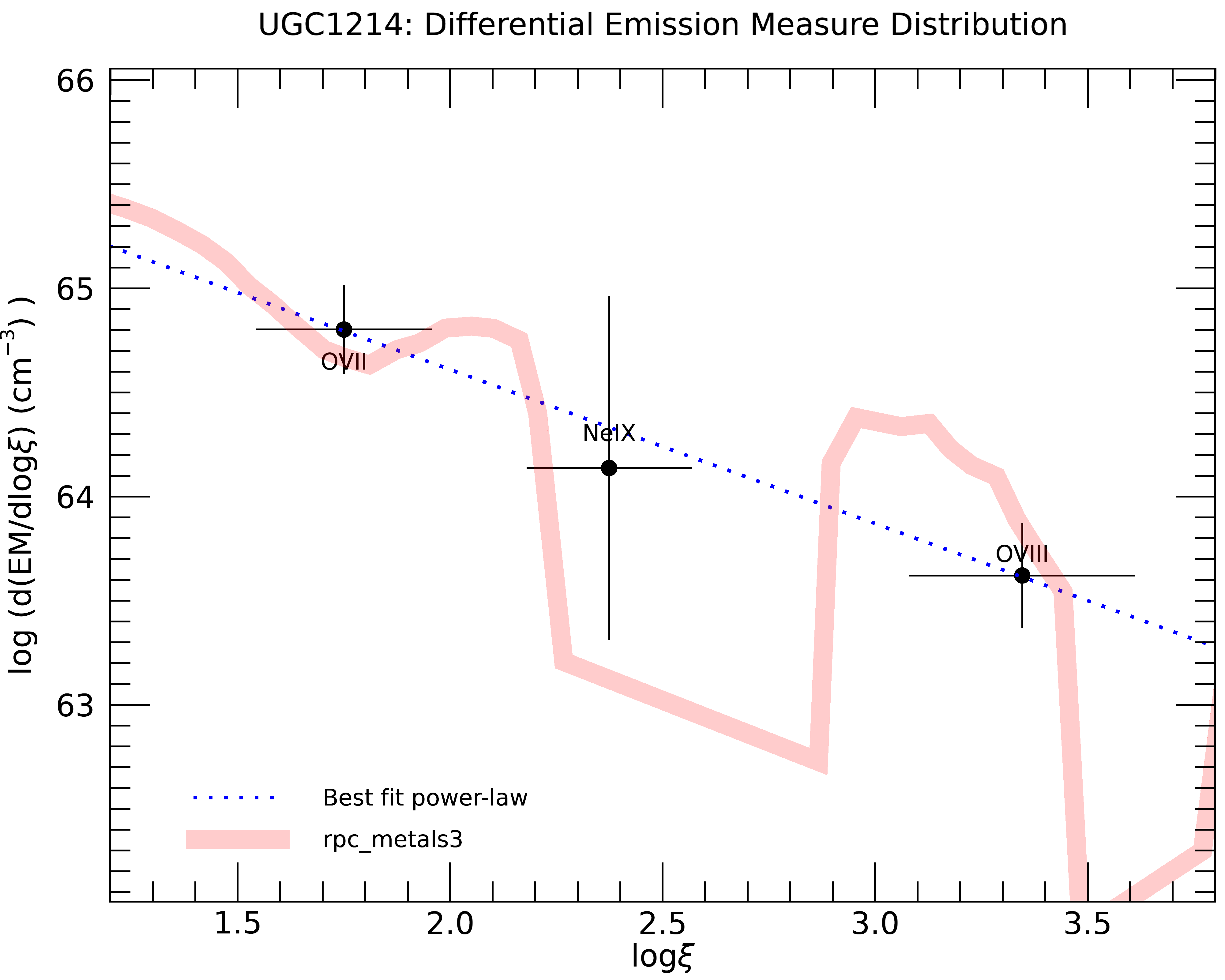}
\caption{\label{EMD5}Same as Fig.~\ref{EMD1}, but for the sources with 3 detected lines: see Table~\ref{tab2}.}
\end{figure*}

\begin{figure}
\includegraphics[width=\columnwidth]{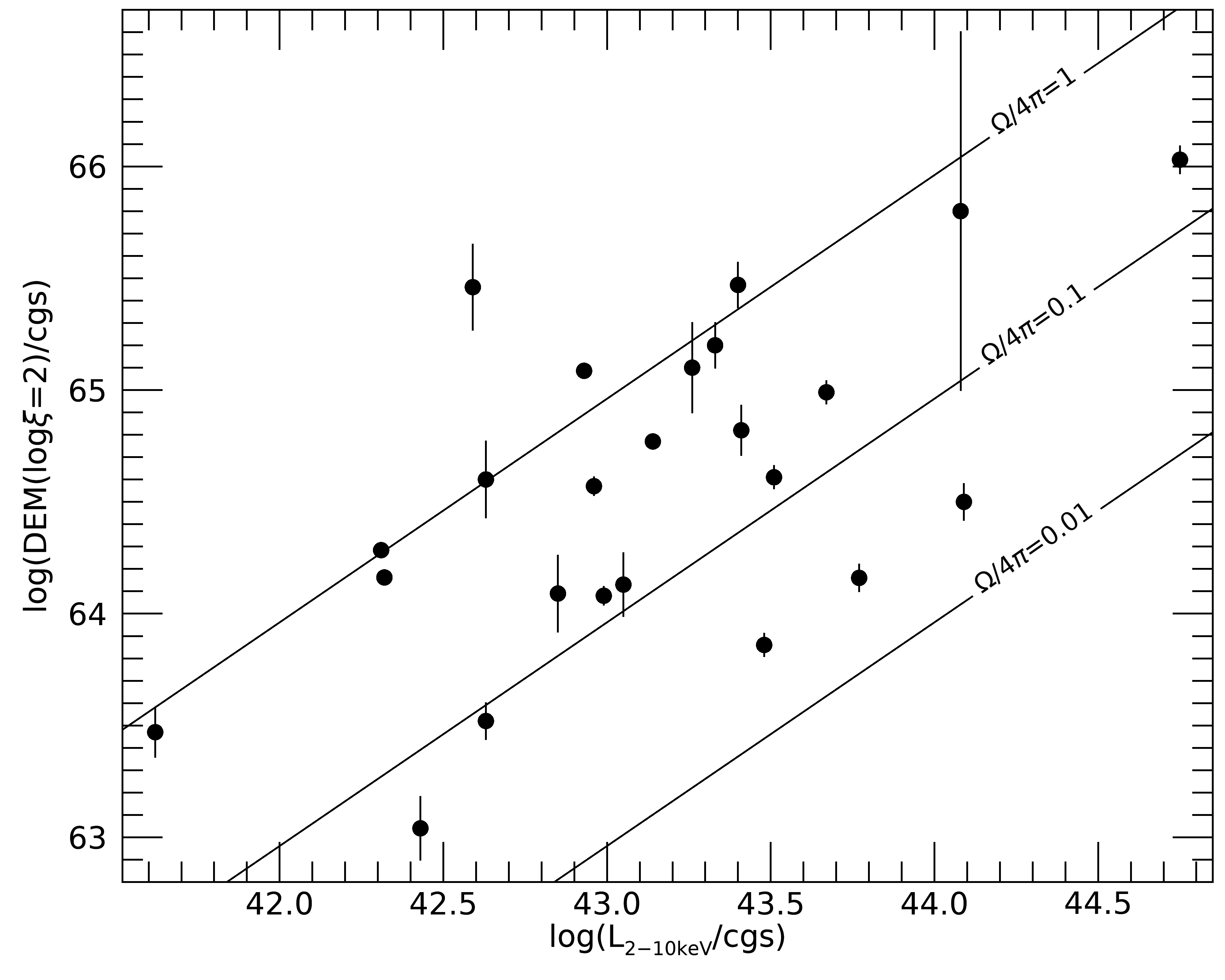}
\caption{\label{CHRESOS_cfactor}Normalization of the observed DEMs at $\log \xi/\mathrm{erg\,cm\, s^{-1}}=2$, assuming the RPC slope vs. the 2-10 keV intrinsic X-ray luminosity of the CHRESOS sources. The solid diagonal lines show the corresponding covering factors of the emitting gas for $\Omega/4\pi=1$, 0.1 and 0.01. See Table~\ref{tab2} and text for details.}
\end{figure}

All the results are summarised in Table~\ref{tab2}, which is also grouped by numbers of detected lines. Along with the best-fit slope for a power-law distribution and the corresponding RPC DEM that best reproduces the data, we also report the normalization of the DEM at $\log \xi/\mathrm{erg\,cm\, s^{-1}}=2$, assuming the RPC slope of $-1.19$ derived above adopting the same 12-point distribution as the data. With this normalization, we can estimate a covering factor for the emitting gas, once compared with the unabsorbed intrinsic 2-10 keV luminosity of each source. This is possible by deriving the predicted ratio between the DEM normalization and the luminosity in \textsc{Cloudy}, where we assumed a covering factor $\Omega/4\pi=1$ (an approximation of this calculation is given in eqn.~\ref{e:DEM RPC num}). The resulting covering factors are reported in Table~\ref{tab2} and shown in Fig.~\ref{CHRESOS_cfactor}, where the normalization of the observed DEMs at $\log \xi/\mathrm{erg\,cm\, s^{-1}}=2$ is plotted against the 2-10 keV intrinsic X-ray luminosity of the CHRESOS sources, together with the derived covering factors.

\section{Discussion}
\label{discussion}

In the previous Sections, we have shown that the DEM predicted by a RPC gas has a universal shape, weakly dependent on the gas properties and the illuminating radiation field (within the parameter space observed in AGN). Therefore, an observed power-law like DEM, with a slope close to -1, is a clear signature from RPC as the dominant compressing mechanism. Indeed, the observed DEMs of the soft X-ray emission in local obscured Seyfert galaxies nicely confirm these predictions, both those derived from the higher S/N spectra and those with less statistics.

A seemingly surprising consequence of these results is that a single high column density ($\log \mathrm{N_H/cm^{-2}}=23$) cloud is capable of reproducing all the observed emission lines. Indeed, there are of course many such clouds, located at a range of distances, but they all produce a very similar emission line spectrum, if characterised by RPC. What is less trivial is that all these clouds have such a high column density. Clouds with lower column densities would contribute only to the higher ionization lines, since they would lack the deeper, less ionised, layer. Their contribution would lead to a flatter DEM distribution with respect to the RPC prediction, as already noted in Sect.~\ref{rpcdem}. This could be the explanation for the few objects in CHRESOS displaying significantly flatter DEM. However, in the majority of the cases, thicker clouds seem to constitute most of the NLR. A likely physical reason for that is that thinner gas filaments may be quickly pushed away, and torn apart by the differential radiation pressure. What survives is only massive gas with high column density, held by the BH gravity. The incident radiation just compresses these gas clouds, leading to a unique and well defined density and ionization structure, supported by the high column density on their back. The edges of the clouds may be still torn apart and pushed away, but this is likely a small fraction of the total line-emitting gas.

In any case, a distribution of clouds with different column densities can only flatten the DEM distribution. To get a steeper distribution, one needs instead clouds which emit only the lower ionization lines, those produced in the back, compressed, side of the cloud. RPC ensures that such a cloud must have the overlying higher ionization surface layers, and so it must produce the whole RPC DEM. The only way to avoid this layer is by having another process which compresses the gas in a totally different way, for example by a shocked-wind bubble of hot gas \citep[e.g.][]{Faucher-Giguere2012}. It is striking to note that no object in CHRESOS has a steeper DEM with respect to RPC. This is further evidence in favour of RPC as the main compressing mechanism in the NLR.

The observed DEMs immediately exclude the much steeper distributions expected for constant density slabs. However, such models are already ruled out by the very different physical scales predicted for emission lines at different ionization, in striking disagreement with the observations. Magnetohydrodynamic (MHD) accretion-disk winds \citep[e.g.][]{Fukumura2010,Fukumura2010a} give an ionization structure that can mimic the observed DEMs, but again predict a stratification of gas at different ionization. The only alternative scenario is therefore a multi-phase medium, where several gas phases, characterized by different parameters, have the same gas pressure, allowing them to be in equilibrium with each other at the same radius. A multi-phase medium does not predict a universal DEM distribution, because the only prescription is that all the phases are in pressure equilibrium with each other, but the relative fraction of each phase is arbitrary. In other words, the observed DEM distributions do not exclude a multi-phase medium, but an arbitrary free parameter is needed in order to reproduce the slope, while the latter is fixed by RPC, whose well-defined structure and distribution of the gas is the only way to counterbalance radiation pressure.

Moreover, only gas pressure is taken into account in the multi-phase scenario, radiation pressure is ignored. But radiation must exercise some pressure once it ionizes and heats the gas, and this cannot be neglected, as shown in Sec.~\ref{introduction} (and references therein). So the gas must have a gradient of (gas) pressure, otherwise it would be necessarily pushed away by radiation pressure (unless radiation is weak, but in this case it would not be ionized at all). In any case, the distinctive feature of a multi-phase medium is that all phases have the same pressure, so directly measuring a gas pressure gradient (for example trough density-sensitive emission lines) would definitely rule out this scenario. On the other hand, extending the DEMs distributions derived here to lower ionization parameters, including Fe-K fluorescence emission and the optical regime from the deepest parts of the cloud, would give further means of testing RPC predictions.

While the overall power-law shape of the DEM predicted for RPC is manifestly recovered in the observed data, it is also clear that the DEM distributions in CHRESOS do not show the characteristic unstable troughs present in the \textsc{Cloudy} computations described in Sect.~\ref{rpcdem}, and likely due to thermal instabilities of the gas \citep[see e.g.][for the similar unstable troughs observed in the absorption measure distributions of warm absorbers]{Holczer2007,Behar2009,Adhikari2015,Goosmann2016a}. On one hand, these unstable troughs are certainly washed away by the rough `resolution' in ionization parameter achieved with the DEM derivation method we use for the observed spectra. Indeed, this effect is already evident in Fig.~\ref{EMD_RPC_lines}: the trough almost disappears in the DEM derived from the emission lines, although it is based on the same \textsc{Cloudy} computation. On the other hand, weaker unstable troughs in the DEM may be encoding an intrinsic physical origin. For example, we have shown in Sect.~\ref{rpcdem} that lower metallicity or steeper X-ray illuminating SEDs yield less pronounced unstable troughs (and indeed the \texttt{ax11} SED is the one that best fits most of the high S/N observed DEMs in CHRESOS). In any case, the precise position and depth of the unstable troughs are sensitive to the micro-physics, and so the details of the troughs need to be treated with caution. Future high-throughput/high-resolution X-ray spectrographs and calorimeters \citep[e.g. \textit{Athena}, \textit{XRISM}, \textit{Arcus}:][]{Nandra2013,Kaastra2017a,Guainazzi2018} will allow for   a much better determination of the DEM, based on more emission lines and in a larger sample of objects, together with accurate estimates of the gas parameters which affect its thermal stability (e.g. density, turbulence).

Equation~(\ref{e:l xi num}) suggests there is a maximum distance beyond which the ionization scale length in RPC is larger than $r$, and hence the plane parallel approximation cannot be applied. This maximum radius can be estimated by substituting $|d r/d \log \xi|$ with $r$, which gives
\begin{equation}\label{e: rmax}
 r_{\rm max} = 47 L_{X; 43} \xi_{100}^{-1.1}\, {\rm pc} 
\end{equation}
This limit depends on $\xi$, so one can equivalently define a maximum $\xi_{\rm max}$ for each $r$. We expect RPC clouds at some radius to have only the layers with $\xi < \xi_{\rm max}$. 
Indeed, the soft X-ray emission in local Seyfert galaxies is extended up to 100s of pc, thus in rough agreement with the adopted plane parallel approximation \citep[e.g.][]{Bianchi2006}.
However, this limit is not unique to RPC, since a similar limit can be derived from the requirement that the line-emitting volume is not larger than the total volume. That is, the DEM cannot be larger than 
\begin{equation}
\mathrm{DEM} = \nH^2 V < \frac{4\pi r^3}{3}\nH^2 
\end{equation}
so using the definition of $\xi$ (eqn.~\ref{eq: xi}) we get 
\begin{equation}
r < \frac{4 \pi L^2}{3 \xi^2 \mathrm{DEM}}  
\end{equation}
Plugging the observed $\mathrm{DEM}$ of NGC~1068 ($10^{65.1}\xi_{100}^{-0.85}\,{\rm cm}^{-3}$, Table~\ref{tab2}) we get a maximum radius of 
\begin{equation}
 r_{\rm max} = 40 L^2_{X;43}\xi_{100}^{-1.15}\, {\rm pc} ~,
\end{equation}
similar to eqn.~(\ref{e: rmax}). 

A special mention is needed for the interesting case of NGC~5548. Historically well-known as an archetypal Type 1 Seyfert galaxy at all wavelengths, NGC~5548 has been found in a persistent X-ray obscured state since at least 2012 \citep{Kaastra2014a}. Its soft X-ray spectrum, now that the intrinsic primary emission is strongly suppressed by the intervening gas, is dominated by emission lines, as in classical obscured Seyfert galaxies \citep{Whewell2015}. We have shown that the derived DEM distribution is also indistinguishable from those observed in the other sources included in CHRESOS, and, in particular, is in perfect agreement with the RPC predictions. In this respect, this represents a lucky (and otherwise impossible) view on the X-ray NLR of a Type 1 object, and a further confirmation of the zeroth-order Unification Model \citep{Antonucci1993}.

\medskip 

Finally, the comparison between the observed DEM distributions in CHRESOS and the RPC predictions allowed us to estimate the covering factors of the X-ray NLR in local Seyfert galaxies. The order-unity covering factors derived in most objects are consistent with derivations of the torus and NLR covering factors derived using other techniques \citep[e.g.][]{Netzer1993,Maiolino2007,Gallagher2007,Treister2008,Stern2012,Lusso2013,Stern2014b}. Some sources have estimates just above unity, but this could be due to to an underestimation of their intrinsic luminosity, which is unobservable and thus strongly model-dependent in Compton-thick objects, like NGC~1068. The only exception is the very large covering factor derived for Mrk~231. However, this is not surprising, since this source is well known to be extremely X-ray weak with respect to its bolometric luminosity \citep{Teng2014}. We do not find any significant correlation between these covering factors and the intrinsic luminosity of the sources. Such a correlation would be instead expected, if the covering factor of the NLR is linked to that of the obscuring/reflecting medium, which is indeed found to decrease with luminosity \citep[see e.g.][for a review]{Bianchi2012a}. However, although the statistical errors are small in most cases, the systematic uncertainties due to the adopted method are quite large and difficult to quantify, and may be the principal source of the observed spread in Fig.~\ref{CHRESOS_cfactor}, as well as wash away any underlying correlation, if present. 

\section{Conclusions}

In an optically thick cloud illuminated by a source of photons, a gas pressure gradient must arise to counteract the incident ionising radiation pressure. The resulting Radiation Pressure Compression leads to a well-defined density and ionization distribution, and thus to a universal Differential Emission Measure (DEM) distribution, with a slope of $\sim-0.9$, which we show is weakly dependent on the illuminating radiation field and the gas properties. 

The observed soft X-ray DEMs of a large sample of obscured AGN with XMM-\textit{Newton} RGS spectra (the CHRESOS sample) are in remarkable agreement with the predicted universal DEM for RPC. Together with the observed spatial and kinematic overlap between soft X-ray emission and the NLR, this provides a clear signature that RPC is the dominant mechanism which sets the gas density, rather than other gas confining mechanisms, such as magnetic fields or the local cloud self-gravity. A constant gas pressure multiphase medium is not ruled out by these results, although it is based on the assumption that radiation pressure is negligible, which is not true. Moreover, the relative fraction of each phase is arbitrary in the constant gas pressure multiphase scenario, so the universal slope of the observed DEMs is not a natural consequence as in RPC. RPC further predicts an increasing gas pressure with decreasing ionization, which can be tested with future high-throughput and high-resolution X-ray micro-calorimeters, using density diagnostics.

In our analysis, we limited the ionization range of the observed DEMs to match the spectral range covered by the XMM-\textit{Newton} RGS, allowing for a uniform analysis of all the sources in CHRESOS. However, the ionization structure of RPC is expected to extend both to higher and lower ionization parameters. We therefore defer to a future work a comprehensive test of RPC predictions at the higher end of the DEM, using H-like and He-like iron recombination lines, and at the lower end, with optical emission lines (e.g. [\ion{O}{iii}]$\lambda5007$) and the iron K$\alpha$ fluorescence line. 

\section*{Acknowledgements}

We would like to thank Ski Antonucci, who triggered this work, and the referee, Sergei Dyda, for his constructive report. SB acknowledges financial support from the Italian Space Agency under grant ASI-INAF 2017-14-H.O. AL acknowledges support by the Israel Science 
Foundation (grant no. 1561/13). JS is supported as a CIERA Fellow by the CIERA Postdoctoral Fellowship Program (Center for Interdisciplinary Exploration and Research in Astrophysics, Northwestern University).

\bibliographystyle{mnras}
\bibliography{RPC}

\label{lastpage}

\end{document}